%% file: ms.tex
\documentclass[preprint2]{aastex}
\usepackage{natbib}
\usepackage{color}
\usepackage{amsmath}
\usepackage{nccmath}

\slugcomment{4 December 2015 Version}

\shorttitle{BESS-POLAR PROTON AND HELIUM SPECTRA}
\shortauthors{K.Abe et al.}

\begin{document}


\title{ Measurements of cosmic-ray proton and helium spectra \\
        from the BESS-Polar long-duration balloon flights over Antarctica}




\author{ K.\thinspace Abe\altaffilmark{1,10}, 
         H.\thinspace Fuke\altaffilmark{2}, 
         S.\thinspace Haino\altaffilmark{3,11}, 
         T.\thinspace Hams\altaffilmark{4,12}, 
         M.\thinspace Hasegawa\altaffilmark{3}, 
         A.\thinspace Horikoshi\altaffilmark{3}, 
         A.\thinspace Itazaki\altaffilmark{1}, 
         K.\thinspace C.\thinspace Kim\altaffilmark{5}, 
         T.\thinspace Kumazawa\altaffilmark{3}, 
         A.\thinspace Kusumoto\altaffilmark{1}, 
         M.\thinspace H.\thinspace Lee\altaffilmark{5}, 
         Y.\thinspace Makida\altaffilmark{3}, 
         S.\thinspace Matsuda\altaffilmark{3}, 
         Y.\thinspace Matsukawa\altaffilmark{1}, 
         K.\thinspace Matsumoto\altaffilmark{3}, 
         J.\thinspace W.\thinspace Mitchell\altaffilmark{4}, 
         Z.\thinspace Myers\altaffilmark{5}, 
         J.\thinspace Nishimura\altaffilmark{6},
         M.\thinspace Nozaki\altaffilmark{3}, 
         R.\thinspace Orito\altaffilmark{1,13}, 
         J.\thinspace F.\thinspace Ormes\altaffilmark{7}, 
         N.\thinspace Picot-Clemente\altaffilmark{5}, 
         K.\thinspace Sakai\altaffilmark{4,9,12}, 
         M.\thinspace Sasaki\altaffilmark{4,14}, 
         E.\thinspace S.\thinspace Seo\altaffilmark{5}, 
         Y.\thinspace Shikaze\altaffilmark{1}, 
         R.\thinspace Shinoda\altaffilmark{6}, 
         R.\thinspace E.\thinspace Streitmatter\altaffilmark{4}, 
         J.\thinspace Suzuki\altaffilmark{3}, 
         Y.\thinspace Takasugi\altaffilmark{1},
         K.\thinspace Takeuchi\altaffilmark{1}, 
         K.\thinspace Tanaka\altaffilmark{3}, 
         N.\thinspace Thakur\altaffilmark{4}, 
         T.\thinspace Yamagami\altaffilmark{2,15}, 
         A.\thinspace Yamamoto\altaffilmark{3}, 
         T.\thinspace Yoshida\altaffilmark{2} and 
         K.\thinspace Yoshimura\altaffilmark{8}}

\affil{$^{1}$ Kobe University, Kobe, Hyogo 657-8501, Japan}
\affil{$^{2}$ Institute of Space and Astronautical Science, Japan Aerospace Exploration Agency (ISAS/JAXA), Sagamihara, Kanagawa 252-5210, Japan}
\affil{$^{3}$ High Energy Accelerator Research Organization (KEK), Tsukuba, Ibaraki 305-0801, Japan}
\affil{$^{4}$ NASA-Goddard Space Flight Center (NASA-GSFC), Greenbelt,MD 20771, USA}
\affil{$^{5}$ IPST, University of Maryland, College Park, MD 20742, USA}
\affil{$^{6}$ The University of Tokyo, Bunkyo, Tokyo 113-0033, Japan}
\affil{$^{7}$ University of Denver, Denver, CO 80208, USA}
\affil{$^{8}$ Okayama University, Okayama, Okayama 700-0082, Japan}

\altaffiltext{9}{Corresponding author.\\
                  \hspace{0.45cm} E-mail address: Kenichi.Sakai@nasa.gov (K.\thinspace Sakai).}
\altaffiltext{10}{Present address: Kamioka Observatory, ICRR, The University of Tokyo, Hida, Gifu 506-1205, Japan.}
\altaffiltext{11}{Present address: Institute of Physics, Academia Sinica, Nankang, Taipei 11529, Taiwan.}
\altaffiltext{12}{Also at: CRESST, University of Maryland, Baltimore County, MD 21250, USA.}
\altaffiltext{13}{Present address: Tokushima University, Tokushima, Tokushima 770-8502, Japan.}
\altaffiltext{14}{Also at: CRESST, University of Maryland, College Park, MD 20742, USA.}
\altaffiltext{15}{Deceased.}


\begin{abstract}
The BESS-Polar Collaboration measured the energy spectra of cosmic-ray protons and helium during two long-duration balloon flights over Antarctica in December 2004 and December 2007, at substantially different levels of solar modulation. Proton and helium spectra probe 
the origin and propagation history of cosmic rays in the galaxy, and are essential to calculations of the expected spectra of 
cosmic-ray antiprotons, positrons, and electrons from interactions of primary cosmic-ray nuclei
with the interstellar gas, and to calculations of atmospheric muons and neutrinos.
We report absolute spectra at the top of the atmosphere for cosmic-ray protons in the kinetic energy range 0.2-160 GeV and helium nuclei 0.15-80 GeV/nucleon. The corresponding magnetic rigidity ranges are 0.6-160 GV for protons and 1.1-160 GV for helium. These spectra are compared to measurements from previous BESS flights and from ATIC-2, PAMELA, and AMS-02. We also report the ratio of the proton and helium fluxes from 1.1 GV to 160 GV and compare to ratios from PAMELA and AMS-02.
\end{abstract}


\keywords{astroparticle physics --- cosmic rays} 

\section{Introduction}

\begin{figure}[!t]
 \centering
 \includegraphics[width=0.5\textwidth]{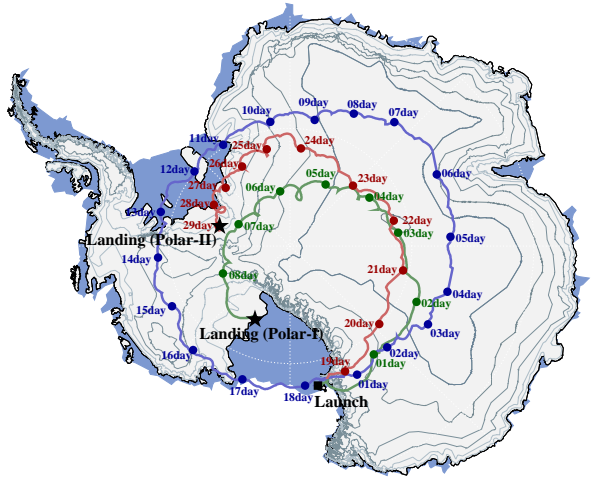}
 \caption{Flight trajectories over Antarctica of BESS-Polar I in 2004 (green) and BESS-Polar II in 2007/2008 (first orbit blue, second orbit red).}
 \label{Fig:Traj}
\end{figure}

We report new measurements of the energy spectra of cosmic-ray protons and helium made by the BESS-Polar Collaboration (Balloon-borne Experiment with
a Superconducting Spectrometer - Polar) during long-duration balloon (LDB) flights over Antarctica in December 2004 (BESS-Polar I), prior to the last solar minimum, and in December 2007 (BESS-Polar II), at solar minimum. The trajectories of both flights are shown in Figure \ref{Fig:Traj}. The absolute fluxes, corrected to the top of the atmosphere (TOA), are reported in energy ranges of 0.2-160 GeV for protons and 0.15-80 GeV/nucleon for helium. In magnetic rigidity ($R = pc/Ze$, momentum divided by electric charge), the corresponding ranges are 0.6-160 GV for protons and 1.1-160 GV for helium. These new spectra are compared to results previously reported by BESS and to measurements from ATIC-2 \citep{Panov2007}, PAMELA \citep{Adriani01042011,Adriani2014323}, and AMS-02 \citep{PhysRevLett.114.171103,PhysRevLett.115.211101}. We combine our measurements to give the ratio of the proton and helium fluxes from 1.1 GV to 160 GV and compare this to ratios reported by PAMELA and AMS-02.

The BESS-Polar measurements reported here are very important to a wide range of astrophysical studies and calculations. Galactic cosmic rays (GCR) near the peak of the all-particle energy spectrum at Earth, $\sim$2 GeV, are made up of $\sim$98$\%$ atomic nuclei (including antiprotons) together with $\sim$2$\%$ electrons and positrons, and the fraction of atomic nuclei increases rapidly with increasing energy. At energies where direct measurements are possible, to about the spectral break known as the ``knee'' at $\sim$3 PeV, GCR nuclei are $\sim$87$\%$ protons and $\sim$12$\%$ helium. The absolute fluxes and spectral shapes of primary cosmic-ray species are central to understanding the origin and the propagation history of cosmic-rays in the galaxy and provide crucial tests of cosmic-ray models.
Proton and helium spectra are essential inputs to calculations of the spectra 
of cosmic-ray antiprotons, positrons and electrons produced as secondary products of 
primary GCR interactions with the interstellar gas. They are also required for calculations of the fluxes of atmospheric muons and neutrinos that are important backgrounds to ground-based neutrino experiments and result from a decay chain beginning with the production of charged pions by interactions of GCR and atmospheric nuclei. 

\section{The BESS-Polar Program}

The BESS instrument \citep{Ajima200071,Haino2004167} was developed as a high-resolution, high-geometric-acceptance
magnetic-rigidity spectrometer for sensitive measurements of cosmic-ray antiparticles, searches for antinuclei, and precise measurements
of the absolute fluxes of light GCR elements and isotopes. Between 1993 and 2002, versions of the original instrument incorporating continuous improvements in performance flew 8 times from Lynn Lake, Manitoba, Canada and once from Fort Sumner, New Mexico, USA   
\citep{Mitchell200431}.

BESS-Polar \citep{Yamamoto20021253,Yoshida20041755,Yamamoto201392,Yamamoto2013227,Mitchell2013} is the most advanced version, based on the general BESS instrument architecture but developed specifically for LDB flights over Antarctica to provide much higher statistics at consistently low geomagnetic cutoff than could be obtained in conventional balloon flights. BESS-Polar presents 45$\%$ less material to incident particles penetrating the full instrument because it has no outer pressure vessel, a thinner magnet coil, and thinner detectors. For the measurements reported here, this lowers the threshold energy to $\sim$200 MeV for protons and $\sim$150 MeV/nucleon for helium. Completely revised low-power electronics support extended LDB flights powered by photovoltaic (PV) arrays and measure the abundant protons and helium nuclei without the event selection required in BESS to control trigger rates.
BESS-Polar I was launched from Williams Field near McMurdo Station, Antarctica,
on December 13th, 2004 and flew for over 8.5 days with the magnet energized at 37 km to 39.5 km (average atmospheric overburden 4.33 g/cm$^2$) and geomagnetic cutoff $R$ below 0.2 GV, recording 2.14 terabytes of data on $9 \times 10^{8}$ cosmic-ray events \citep{Abe2008103}.

BESS-Polar II incorporated considerable improvements in instrument and payload systems, including improved magnet cryogen hold time, reduced deadtime per event, and much greater onboard storage capacity. It was launched on December 23, 2007 and observed cosmic rays for 24.5 days with the magnet energized at 34 km to 38 km (average of 5.81 g/cm$^{2}$) and cutoff $R$ below 0.5 GV, accumulating 13.6 terabytes of data on $4.7 \times 10^9$ events \citep{Mitchell2013}.

Previous papers from the BESS-Polar program have reported precise measurements of  
antiprotons \citep{Abe2008103, PhysRevLett.108.051102} and a sensitive antihelium search
 \citep{PhysRevLett.108.131301} and discuss instrumentation specific to those measurements. 

\section{The BESS-Polar Instruments}

\begin{figure*}[!t]
 \centering
 \includegraphics[width=0.9\textwidth]{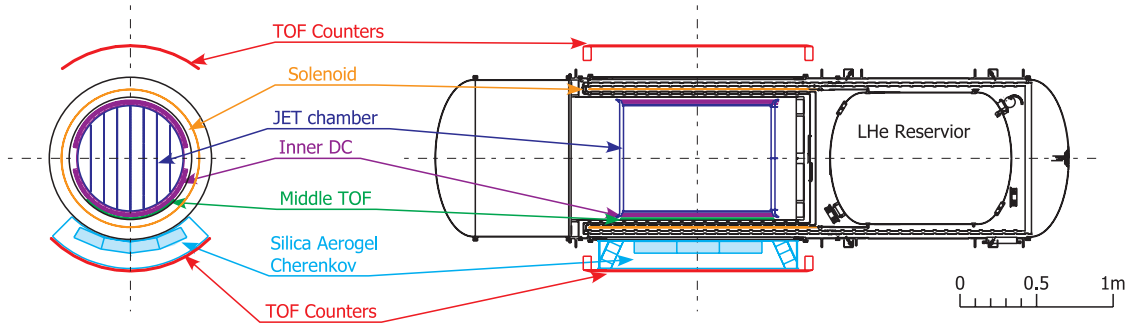}
 \caption{Crosssection and side views of the BESS-Polar II instrument}
 \label{Fig:Det}
\end{figure*}

In both BESS-Polar instruments (Figure \ref{Fig:Det}), 
a uniform magnetic field is produced by a thin superconducting solenoid magnet 0.9 m in diameter and 1.4 m long, incorporating a 0.8 m diameter, 1.4 m long coaxial warm bore in which detectors are located. The magnet is capable of stable operation at 1 Tesla (1 T), but in both flights was operated at 0.8 T for additional stability margin. Drift-chamber tracking detectors in the warm bore measure $R$. The cylindrical configuration of the BESS-Polar spectrometer gives it a large angular acceptance. With $\sim$1 $m^2$ time-of-flight (TOF) hodoscopes separated by $\sim$1.6 m, the nominal geometric acceptance of BESS-Polar is $\sim$0.3 m$^2$sr at 10 GV. 

The primary TOF system composed of upper (UTOF) and lower (LTOF) plastic scintillator hodoscopes measures velocity ($\beta$)
and differential energy loss ($dE/dx$) and forms the event trigger. A thin plastic scintillator middle-TOF (MTOF) hodoscope is installed on the lower surface of the magnet bore to measure low-energy particles that cannot reach the LTOF. The MTOF further lowers the threshold energy to about 100 MeV for antiproton or proton measurements, but has not been used here because the low energy spectra are strongly affected by solar modulation and our focus is on higher energies.

A threshold-type silica-aerogel Cherenkov counter (ACC), below the spectrometer, helps reject $e^{-}$ and $\mu^{-}$
backgrounds to the antiproton measurements. This is not used in the proton and helium analysis reported here because contamination by light backgrounds is insignificant. Its use in the antiproton analysis has been described previously \citep{Abe2008103, PhysRevLett.108.051102}.

Particle electric charge ($Ze$) is determined using $dE/dx$ and $\beta$. Particle mass ($m$) is calculated from $R$, $Z$, and $\beta$ as: 
 \begin{equation}
  m = RZe{\sqrt{1/\beta{^2}-1}} .
 \end{equation}
Charge sign is determined by the direction of curvature in the magnetic field.

The superconducting magnet uses NbTi/Cu wire with high-strength aluminum stabilizer incorporating Ni doped filaments to increase its strength \citep{Makida2009}. The magnet and cryostat have only 2.46 g/cm$^{2}$/wall. The magnet operates in persistent mode with a cryogen hold time of 11 days in BESS-Polar I, using a 400-liter liquid helium (LHe) reservoir, improved to 25 days in BESS-Polar II with 520 liters of LHe. The tracking system uses a JET-type drift chamber (JET) with essentially no material in the tracking volume, other than gas, to minimize scattering. Inner drift chambers (IDCs) with two tracking layers each are located above and below the JET. 

The JET and IDCs accurately determine the position of particles passing through each drift cell from the recorded drift times and measure their $dE/dx$ from the digitized signal amplitudes at the sense wires. Pure CO$_{2}$ fills the drift chambers and fresh gas is circulated during flight by a semi-active flow control system and exhausted through the pressurized warm-bore of the magnet to ensure that the drift gas is not contaminated. The signals of the JET and IDCs are read out using flash analog-to-digital converters (FADC) capable of recording both the time of detection at the sense wire and the shape and amplitude of the detected pulse. Particle tracks are measured by up to 52 points (48 from the JET and 2 from each of the 2 IDCs) with $\sim$125 $\mu$m resolution in the bending (r$\phi$) direction (perpendicular to the magnetic field). The uniform magnetic field of the BESS-Polar solenoid makes the $R$ measurements relatively insensitive to tracking accuracy in the $z$ coordinate (parallel to the magnetic field), and this is accomplished using vernier pads in each layer of the IDCs, giving $\sim$650 $\mu$m resolution, and JET measurements using charge division to give $\sim$24 mm resolution. The continuous tracking with minimal scattering provided by its tracking system enables BESS-Polar to readily distinguish interactions and hard scatters. A truncated mean of the integrated charges measured by the JET provides an energy loss ($dE/dx$) measurements with a resolution of $10\%$.
The maximum detectable rigidity (MDR, the highest $R$ curved track distinguishable from a straight line) is 240 GV for the longest tracks through the spectrometer, expecting 52 points to be used in fitting tracks. 

The scintillator hodoscopes consist of 10 UTOF and 12 LTOF Eljen E204 scintillator paddles, each 0.95 m long with cross sections 96.5 mm wide $\times$ 10 mm thick in BESS-Polar I and 100 mm $\times$ 12 mm in BESS-Polar II. Hamamatsu R6504 magnetic-field-tolerant, fine-mesh photomultiplier tubes (PMTs) at each end view the scintillators through UV-transmitting acrylic light guides. The timing resolution is 120 ps, giving a $\beta^{-1}$ resolution of 2.5\%. The nominal BESS-Polar event trigger is a four-fold coincidence that requires a PMT on each end of both the UTOF and LTOF to exceed a fixed signal threshold. Coincidences of the UTOF and MTOF also trigger events.

TOF signals are digitized by low-power common-stop time-to-digital converters (TDCs) attached to the PMT anodes and 16-bit charge-to-digital converters (QDCs) attached to both the 13th and 18th dynodes to increase dynamic range. The 18th dynode measurements are normally used to measure $dE/dx$ for light particles and nuclei. As noted, JET and IDC signals are digitized by FADCs. All readout electronics are highly parallel and the full processing and communication time for each event is $\sim$40 $\mu$s. Event data are recorded to onboard hard disk drives with a total capacity of 3.5 terabytes (TB) in BESS-Polar I and 16 TB in BESS-Polar II. The data acquisition event rate was $\sim$1.4 kHz in BESS-Polar I (due to PMT issues that reduced the acceptance, discussed below) and $\sim$2.5 kHz in BESS Polar II. Total power consumption was $\sim$420 W. Because the magnet fixed the orientation of the payload with respect to the local geomagnetic field, all sides of the payload were exposed to the sun each day and an omnidirectional photovoltaic array was required.

\section{In-Flight Instrument Issues} \label{ssec:ii}

The UTOF and LTOF operate at ambient pressure ($\sim$5 mbar) and protection against high-voltage (HV) breakdown was required. In BESS-Polar I this was accomplished by potting the high-voltage components. In the BESS-Polar I flight, 18 of 44 PMTs in the TOF had to be turned off because their potting failed under a combination of cold and vacuum, and they began to draw excessive current. This required the trigger to be changed in flight to a two-fold coincidence between any UTOF PMT and any LTOF PMT. Although some paddles did not have a working PMT and did not provide trigger signals, this retained 74$\%$ of the nominal trigger geometric acceptance. In BESS-Polar II there were no HV issues due to vacuum, but 2 of the 44 PMTs had HV control problems and were turned
off, one on the UTOF and one on the LTOF. With a change to a two-fold coincidence, all TOF paddles provided trigger signals and 100$\%$ of 
the nominal trigger geometric acceptance was retained. 

In BESS-Polar I, the tracking system was stable throughout the flight. In BESS-Polar II, however, the JET HV suddenly became 
unstable after more than a day at float and the voltage and 
the current began to fluctuate. The HV was immediately reduced to 90$\%$ of 
the nominal operating value and the CO$_{2}$ pressure was reduced to keep the same
drift velocity. This intervention succeeded and the chamber continued to function well for the remainder of the flight, although the HV fluctuated for periods between two distinct voltages. Despite this, normal tracking resolution was obtained for more than 90$\%$ of the science
observation time by development of algorithms that calibrate the tracker over short
time intervals and depend on its HV state. Time lost when the HV was unstable is reflected in the trigger live time discussed in Section \ref{ssec:dq}. Straight tracks obtained 
in flight without magnetic field after the LHe was expended were used as an absolute reference of events with
``infinite $R$" to check the calibration and ensure that it was not biased.

\section{Data Analysis}

\subsection{Tracking System Calibration}

Each track point in the drift chambers was calculated from the measured drift times in the JET drift cells and the IDCs.
The position reconstruction started with 
space-time templates calculated by applying the CERN GARFIELD simulation code \citep{Verrnhof1998} to the nominal design. These templates were adjusted to match the actual chambers by iteratively modifying
the parameters.
The calibration process considered FADC time offsets, electric-field and CO$_{2}$ pressure-dependent drift velocities, Lorentz-angle corrections,
wire position displacements, and field distortion corrections. Tracks through the magnetic field were constructed from the measured points and used to calculate $R$. 

The limitations to the accuracy of 
the $R$ measurements arise from the characteristics of the spectrometer including the integral magnetic field, field uniformity and absolute value, track length in the field, and tracking resolution. There are also potential uncertainties from deflection offsets due to small misalignments of the sense wires. 

Prior to flight, the magnet was energized to a central magnetic field of 0.8 T, and then put into persistent current mode. The subsequent decay of the magnetic field is very slow and it was essentially constant for the duration of the flight. Thus, the uncertainty in the absolute magnetic field is dominated by the uncertainty in the excitation current, estimated as 
$0.5\%$. The resulting systematic uncertainties in the observed fluxes were estimated by varying the magnetic field $\pm0.5\%$ from the nominal value and found to be $<1.5\%$ for both protons and helium.

The measured distribution of the deflection resolution ($\sigma (R^{-1})$) in both BESS-Polar I and BESS-Polar II has mean of 4.2 TV$^{-1}$ for tracks with the longest possible lengths in the spectrometer. Over the full fiducial volume used in this analysis, the mean of the measured deflection distribution is 5.0 TV$^{-1}$ corresponding to an MDR of 200 GV. The precise alignment of the JET with respect to the IDCs was determined by checking the 
consistency between tracks reconstructed by JET and points measured by IDCs. 
Wire position displacements were determined 
using ground muon data for BESS-Polar I and with
data taken without magnetic-field at the end of the flight for BESS-Polar II.
A small net displacement was observed in both cases.
The wire position calibration of the JET gives a maximum sense-wire alignment uncertainty of $\leq$10 $\mu$m. It was expected that this would be random and would have no significant effect on performance. However, a coherent misalignment of all wires by the maximum 10 $\mu$m could introduce a systematic offset in the deflection
measurement of $\Delta R^{-1}$ = 0.33 TV$^{-1}$ and result in the observed deflection displacement.
This was included as part of the systematic error attributed to chamber alignment. A detailed model of the structure, operating conditions, and response function of the drift chamber tracker 
was implemented in a Geant4 Monte Carlo simulation and the simulated $R$ resolution accurately reproduced the measured resolution. 

\subsection{Track Reconstruction and Acceptance} \label{ssec:tf}
Tracks were fit by an iterative method developed to give high track-finding efficiency with little rigidity dependence. An initial track was first generated using all valid chamber hits. This was then refined by removing from the fitting all wires whose measured track position was more than 20 $\sigma_{r\phi}$ away from the corresponding value from the previous track fit and then re-fitting the track using only the surviving points. This process was repeated for 15, 10, and finally 5 $\sigma_{r\phi}$ and the final track was fit using points from those wires that survived all cuts.

For the best $R$ resolution consistent with maximum efficiency and geometric acceptance, particles are required to traverse the tracker completely within a fiducial volume defined by the central six columns of drift cells (out of eight) in the JET and a requirement on the cosine of the zenith angle of
$\cos \theta_{\rm zenith} > 0.80$. This reduced the geometric acceptance at 10 GV to 0.149 m$^2$sr in BESS-Polar I (due to TOF PMT issues discussed in Section \ref{ssec:ii}) and 0.199 m$^2$sr in BESS-Polar II. For both instruments this requirement gives $\geq$40 points expected from the JET and 4 points required from the IDCs. 
The error comes from the uncertainty of the detector alignment of less than 1 mm, corresponding to about $0.2\%$ systematic error of the acceptance by UTOF and LTOF.

\subsection{Monte Carlo Simulated Efficiencies}  \label{ssec:se}
Selection criteria were applied as cuts to the event data. These are characterized by cut selection efficiencies defined as the probability of an event surviving a particular cut. There are two classes of selection. Selections whose effects cannot be determined directly from the flight data are discussed here. Their corresponding selection efficiencies are estimated using Geant4 (Version 10.00.p01) Monte Carlo simulations and are combined into an overall selection efficiency $\varepsilon_{\rm MC}$. Selections whose effects can be directly determined from flight data are discussed in Section \ref{ssec:dq}.

The selection criteria were almost identical in 
BESS-Polar I and BESS-Polar II analyses with the following exceptions. 
In the BESS-Polar I analysis, we allowed only one hit paddle on either the UTOF or LTOF to be read out by a single PMT and required the hit paddle on the other TOF layer to be read out by 2 PMTs. The corresponding selection efficiency was $59\%$ for protons and $62\%$ for helium.
In BESS-Polar II, 2 of the 34 IDC wires were excluded because of low signals due to the HV supply issue. To ensure the best TOF resolution and to allow $z$ position constancy between the TOF and tracking to be used, we conservatively required each TOF paddle to be read out by 2 PMTs. The selection efficiency from the combination of these cuts were $58\%$ for proton and $61\%$ for helium.

To eliminate rare events in which an interaction took place in or near the instrument, single-track events were required having only
one isolated track and no more than two paddles hit in each layer
of the TOF hodoscopes. The non-interaction selection efficiency as a function of $R$ was estimated by applying the single-track criteria to events from Geant4 simulations. At 10 GeV/nucleon it was about $90\%$ for protons and $77\%$ for 
helium. The error in the non-interaction efficiency was estimated 
by comparing the number of LTOF paddles hit in the flight data and in the simulated data.

The resolution of the IDCs deteriorates for tracks passing both near the sense wires and at large drift distances. To ensure the best tracking, an
effective IDC fiducial region of about $\sim$85$\%$ of the active area was defined by cuts on the drift time to exclude the regions very near the sense wires ($\sim$10$\%$ of the drift distance) and far from the wires ($\sim$5$\%$ of the drift distance). 
The resulting selection efficiency was $54\%$ in BESS-Polar I and $56\%$ in BESS-Polar II. 

\subsection{Data Quality Cut Efficiencies} \label{ssec:dq}
The efficiencies of data quality cuts are determined directly from flight data by the ratio of events after and before the cut, unlike the event selections discussed in Section \ref{ssec:se}. In BESS-Polar, these include track quality cuts and electronic hazard flag cuts. The rigidity of each particle is determined with good accuracy prior to these cuts. This is discussed in more detail in Section \ref{ssec:rd}.

To assure the accuracy of the rigidity measurement, the following track quality cuts were
applied: (1) reduced $\chi^2$ per degree of freedom $<$5 for track fits in both $r\phi$ and $yz$ planes (where $y$ is the vertical direction), 
(2) track fitting pathlength $\geq$500 mm, 
(3) ratio of fitted hits in the JET to expected hits $>$0.6,
(4) ratio of dropped hits in the JET and IDCs to expected hits $<$0.25,
(5) 2 hits in each of the 2 IDCs all giving both x and z positions, and (6) consistency between independent $z$ positions measured by the tracking system and the TOF (determined using the time difference between
 each end of the hit paddle). The efficiencies from (1) to (4) were $\sim98\%$ for protons and helium nuclei
in both flights.

At 10 GeV/nucleon, the efficiencies for protons surviving (5) were 
$(89.53 \pm 0.08)\%$ in BESS-Polar I and  
$(90.37 \pm 0.03)\%$ in BESS-Polar II.
The efficiencies from (5) for helium were 
$(72.09 \pm 0.46)\%$ in BESS-Polar I and 
$(71.07 \pm 0.18)\%$ in BESS-Polar II.
Helium efficiencies were almost $20\%$ 
lower than for protons due to misidentification from delta-rays and saturation (overflow) of the FADCs causing
$5 \%$ of IDC hits to be cut. 

At 10 GeV/nucleon, the efficiencies for protons surviving (6) were 
$(97.55 \pm 0.04)\%$ for BESS-Polar I and
$(98.61 \pm 0.01)\%$ for BESS-Polar II.
The efficiencies from (6) for helium were
$(98.26 \pm 0.13)\%$ for BESS-Polar I and 
$(98.65 \pm 0.04)\%$ for BESS-Polar II.  In BESS-Polar I, those detectors with only a single operating PMT did not provide redundant $z$ positions and cut (6) could not be used. 

The TOF QDCs incorporated hazard flags that were set when the QDC was performing a function that might give an incorrect measurement. Events displaying hazard flags were rejected. In BESS-Polar I, the QDCs were operated in a mode in which alternating capacitors were used to sample background noise, which was then subtracted electronically from the measurements. This process proved to be inefficient at flight trigger rates. In BESS-Polar II, the efficiency was greatly improved because the QDCs were operated with a delayed input and no background subtraction was required. The efficiency of surviving this cut was estimated as
$(52.00 \pm 0.09)\%$ for protons and
$(52.23 \pm 0.35)\%$ for helium nuclei at 10 GeV/nucleon in BESS-Polar I, and as
$(99.62 \pm 0.01)\%$ for protons and 
$(99.62 \pm 0.02)\%$ for helium nuclei in BESS-Polar II. 
The trigger live time during which events were recorded was measured exactly by counting 1 MHz clock pulses as
472881 seconds in BESS-Polar I and 
1273381 seconds in BESS-Polar II. Combining the trigger live time and the hazard flag cut efficiency give the actual measurement live time $T_{\rm live}$ used in the flux calculations.

\subsection{Flight Data Estimated Efficiencies}  \label{ssec:de}
The final class of efficiencies that must be considered are due to instrument operation or event analysis, but do not result from event selections. These are estimated from independent flight data selections.

The event trigger efficiencies depend on the detection efficiencies from discrimination of signals in the UTOF and LTOF paddles through which a particle passed and the efficiency of the event trigger logic. These were estimated using particles that triggered coincidences of the UTOF and MTOF, passed through all TOF layers (UTOF, MTOF, LTOF), and had $R >10$ GV. The trigger efficiencies were estimated from the probability that a normal event trigger was also generated. These were $(99.6 \pm 0.1)\%$ in BESS-Polar I and $(99.7 \pm 0.1)\%$ in BESS-Polar II.

The JET chamber has insensitive regions very close to the sense wires. When a particle passes through several such regions near the center of the JET, the track reconstruction algorithm cannot succeed. The corresponding track reconstruction efficiencies were estimated by scanning $5\times10^3$ unbiased events with appropriate TOF hits and enough overall JET points. The efficiencies for both proton and helium were $(99.0 \pm 0.2)\%$ in BESS-Polar I and $(98.7 \pm 0.2)\%$ in BESS-Polar II.

When a particle unrelated to a measured event (an ``accidental") passes through the JET or TOF within the resolving time, the single-track selection discussed in Section \ref{ssec:se} eliminates the event. The influence of such accidental particles can be estimated by utilizing data recorded for random triggers that were issued independent of particle triggers at a rate of 10 Hz throughout the flight. Particles seen in these random trigger events were used to estimate the rates of accidental particles that might have caused real events to fail the single-track selections. The accidental efficiencies were $97.7\%$ in BESS-Polar I and $96.2\%$ in BESS-Polar II.

\subsection {Particle Identification} \label{ssec:pi}

\begin{figure}[!t]
 \centering
 \includegraphics[width=0.50\textwidth]{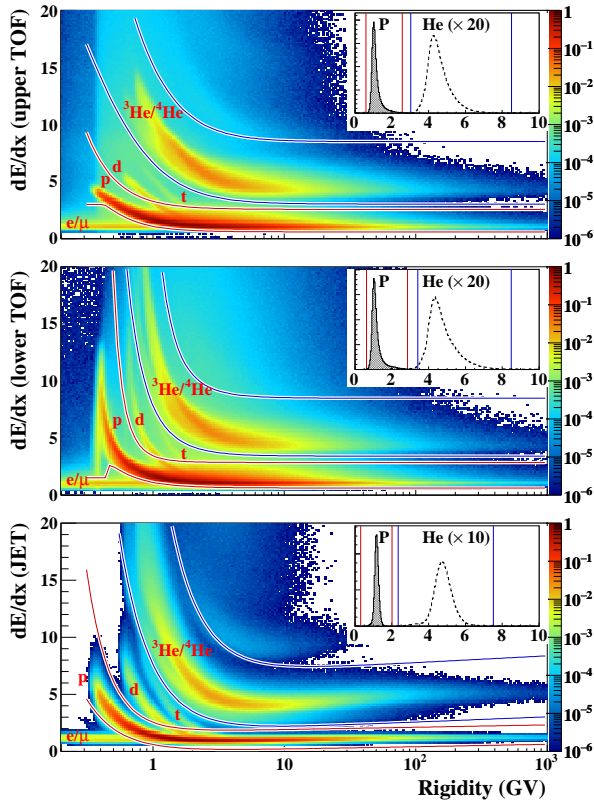}
 \caption{Proton and helium bands in $dE/dx$ vs $R$  (top: UTOF, middle: LTOF, bottom: JET) from BESS-Polar II. The superimposed lines show the selection criteria for protons and helium nuclei. The inset graphs show the results of those selections above 30 GV.}
 \label{Fig:dedx}
\end{figure}

\begin{figure}[!t]
 \centering
 \includegraphics[width=0.50\textwidth]{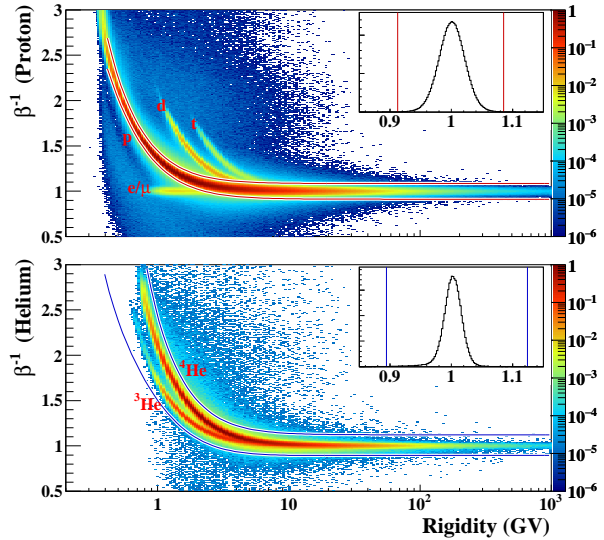}
 \caption{BESS-Polar II proton (top) and helium (bottom) bands in $\beta^{-1}$ vs $R$ after the selections shown in Figure \ref{Fig:dedx}. The superimposed lines show the selection criteria for protons and helium. The inset graphs show the results of those selections above 30 GV.}
 \label{Fig:beta}
\end{figure}

Protons and helium nuclei were identified by $dE/dx$ vs $R$ and $\beta^{-1}$ vs $R$ techniques using the UTOF and LTOF. 
As examples, Figures \ref{Fig:dedx} and \ref{Fig:beta} show the selection
bands for protons and helium in BESS-Polar II.
In BESS-Polar I, protons and helium nuclei were identified in the same manner.
The efficiency of the TOF $dE/dx$ vs $R$ selections were estimated by applying them to an independent sample of events
selected using the JET $dE/dx$ vs $R$.
Overall, $(95.84 \pm 0.50)\%$ of protons and  
$(92.19 \pm 1.50)\%$ of helium nuclei were properly identified in BESS-Polar I.
In BESS-Polar II, 
$(96.09 \pm 1.00)\%$ of protons and  
$(93.39 \pm 1.50)\%$ of helium nuclei were properly identified.
The errors were estimated by comparing the TOF $dE/dx$ efficiencies in the flight data and in the simulated data.

Since the $\beta^{-1} $distribution
is a Gaussian and the half-width
of the $\beta^{-1}$ selection band was set at 4 $\sigma$, the
efficiency is very close to unity for protons.
In order to retain helium nuclei which might have been misidentified due to the contributions of accompanying $\delta$-rays,
the width of the selection band was wider in the helium analysis at 8 $\sigma$. With this adjustment the helium selection efficiency was also close to unity.

\subsection {Measurement Dependence on Rigidity} \label{ssec:rd}
 
The proton and helium measurements made by the BESS-Polar instruments span wide ranges of $R$. Their uniform performance over these wide ranges is demonstrated by the very small dependence on $R$ of two figures-of-merit, the overall detection efficiency and the effective acceptance. They are both nearly constant, especially at higher $R$, giving great confidence in the measured slopes of the interstellar proton and helium spectra. 

The track quality cuts required the positions hits in the tracking detector used in the analysis to fall near the fitted track. Analysis of these cuts at rigidities where multiple scattering can be neglected show that they do not depend on rigidity. Below 100 GV the tracking resolution before tracking quality cuts is adequate to measure the rigidity to a small fraction of a rigidity bin and fluxes with and without quality cuts are essentially identical. There is no migration of events between rigidity bins due to the cuts.

\begin{figure}[!t]
 \centering
 \includegraphics[width=0.5\textwidth]{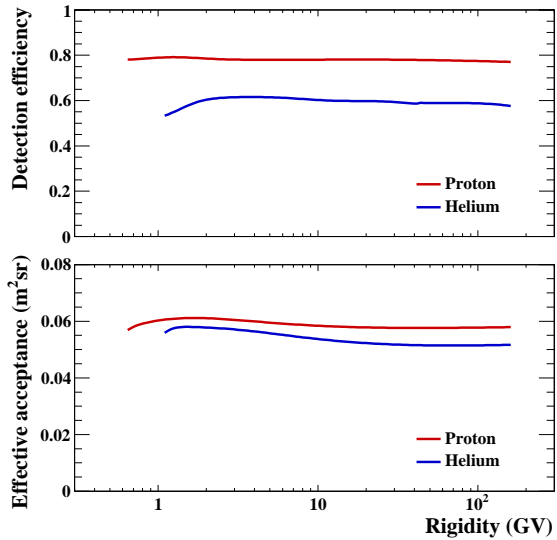}
 \caption{BESS-Polar II detection efficiencies for protons and helium (top) and effective acceptances (bottom) as functions of $R$.}
 \label{Fig:eff}
\end{figure}

Figure \ref{Fig:eff} (top) shows the detection efficiencies, $\varepsilon_{\rm det}$, for protons and helium nuclei 
from the BESS-Polar II analysis, combining the track quality, track reconstruction, trigger, accidental, and particle identification efficiencies.
At the lowest $R$, some signals in the JET overflowed the FADCs due to high $dE/dx$ and inconsistent $z$ tracking between the JET and IDC slightly reduced detection efficiencies for helium. Above $\sim$2 GV the efficiencies for both protons and helium were essentially constant.
The detection efficiencies were stable with time and the efficiencies above 20 GV for 4-hour time intervals are estimated as  
 $(78.62 \pm 0.93)\%$ for protons and $(59.35 \pm 1.88)\%$ for helium nuclei in BESS-Polar I.
In BESS-Polar II, they were $(78.21 \pm 0.60)\%$ for protons and $(59.41 \pm 1.21)\%$ for helium nuclei. 

To illustrate the uniformity of the BESS-Polar particle acceptance over the full range of $R$, we define an effective acceptance as the product of the geometric acceptance ($S\Omega$) and the selection efficiencies discussed in Section \ref{ssec:se}. This is shown in Figure \ref{Fig:eff} (bottom). The effective acceptances at 10 GeV/nucleon in BESS-Polar I are 
0.043 m$^{2}$sr for protons and
0.039 m$^{2}$sr for helium. In BESS-Polar II they are
0.058 m$^{2}$sr for protons and 
0.052 m$^{2}$sr for helium. 
The systematic error in the effective acceptance primarily results from the systematic error in the non-interaction efficiency. 

\subsection{Backgrounds and Measured Species}

\subsubsection{Light Particles} 

Interactions of cosmic rays with the atmosphere above BESS-Polar and with the balloon-flight suspension components, especially the swivel, produce light particles, including positrons, pions, and muons, that form a background to the proton measurements. Below 1.6 GV the protons are clearly separated from the lighter particles, as shown in Figure \ref{Fig:beta}, and so are uncontaminated. However, above 1.6 GV the light particles can no longer be identified.
To estimate the magnitude of the resulting background, the ratio of light particles to protons was measured as a function of rigidity in the range where both can be clearly identified. The light particle to proton ratio was $2.1\%$ at 0.8 GV, rapidly dropping to $0.8\%$ at 1.6 GV, and could be fit well by a power law in $R$ with
an index of $-1.6$, indicating that contamination is negligible above 1.6 GV. The measurements were reproduced well by solving simultaneous transport 
equations using as input BESS-Polar II protons and helium nuclei and electrons and positrons 
from PAMELA measurements. Because both the measurements and calculations showed that contamination of the measured protons by light particles is negligible, no corrections were made.

\subsubsection{Deuterons and Tritons} 

The BESS-Polar spectrometers distinguish deuterons up to 3 GV and our measurements show that the $d/p$ ratio decreases with 
increasing rigidity, falling to $2.5\%$ at 3 GV
in BESS-Polar I. Because of its greater atmospheric overburden, the $d/p$ ratio at 3 GV is $3.0\%$ in BESS-Polar II. Above 3 GV, the deuterons could not be resolved. Similarly, tritons can be resolved up to 5 GV, where the measured $t/p$ ratio was $\sim 0.5\%$. Above 5 GV tritons could not be resolved. Thus, above 3 GV the reported protons have a small admixture of deuterons and above 5 GV an even smaller admixture of tritons.
We conservatively include $\sim$3$\%$ uncertainty for deuteron contamination that may have increased the measured proton flux. Tritons were not included in the uncertainty.

\subsubsection{$^{3}$He}

 In both flights, $Z$=2 nuclei were clearly identified by $dE/dx$ vs $R$ and $\beta^{-1}$ vs $R$ techniques using the UTOF and LTOF hodoscopes and were confirmed using $dE/dx$ vs $R$ from the JET. However, these nuclei included both $^{3}$He and 
$^{4}$He. Following the practice used in previous BESS reports of He fluxes, all $Z$=2 particles were included in the He spectrum.

\subsection{Flux Calculations and Corrections}

\subsubsection{Top of Instrument Flux} 

Nuclei that have enough energy to penetrate to the LTOF and trigger the instrument have nearly constant energy throughout the tracker. Thus, the energy that the particle would have had when entering the instrument (top-of-instrument or TOI) is calculated by integrating its differential energy losses $dE/dx$ for all materials above the tracker using a detailed model of the instrument components encountered along its trajectory. The measured differential flux of a given particle species at TOI ($\Phi_{\rm TOI}$), binned in energy with width d$E$, can then be determined by: 

\begin{equation}
\Phi_{\rm TOI} (E_{\rm TOI})= \frac{N_{\rm p}}{\varepsilon_{\rm det} \cdot \varepsilon_{\rm MC} \cdot S\Omega \cdot T_{\rm live}} ,
\end{equation}

\noindent
where $N_{\rm p}$ is number of observed particles of that species.

\subsubsection{Spectral Deconvolution}

For steeply falling particle spectra, observed spectral shapes can be deformed because of the $R$ dependence of the magnetic-rigidity spectrometer $R$ resolution, or because of the energy dependence of the energy resolution of a calorimeter, extensive air shower detector, etc.. The details of this deformation depend on the absolute $R$ or absolute energy dependence of the measurement resolution. Response functions of different measurement techniques differ with energy and sometimes (e.g. calorimeters) with particle species. For magnetic-rigidity spectrometers like BESS-Polar, the resolution deteriorates as $R$ approaches the MDR. The more abundant lower $R$ particles are misidentified as higher $R$, and, by number conservation, fluxes are correspondingly depleted at slightly lower $R$, giving an apparent dip in the spectrum and then a recovery to a harder spectral index than the underlying spectrum. Depending on the underlying spectral index, small $R$ errors of this sort can result in a considerable change in the observed spectral shape and power law index (or indices). To compensate, reported high $R$ spectra from magnetic rigidity spectrometers are usually the result of a ``deconvolution" procedure to reproduce the most probable real spectrum that would have resulted in the observed spectrum, taking into account the finite $R$ resolution. For the results reported here, the deconvolution was accomplished using a statistical procedure based on iteratively applying Bayes' 
theorem.
In this method, the finite resolution effect is described by a convolution matrix 
in deflection space, which was estimated from a simulation using the appropriate $(R^{-1})$ resolution. As noted above, for both BESS-Polar I and BESS-Polar II, $\sigma (R^{-1})$ = 5.0 TV$^{-1}$.
An initial spectrum must be chosen as the starting point for the deconvolution. Based on results from previous measurements in the BESS-Polar rigidity range, single power laws with spectral indices of $-2.80$ for protons and
$-2.70$ for helium nuclei were used. The systematic error of the deconvolution procedure
was estimated to be about $1\%$ at 160 GV from the deviation due to input index errors
($\pm 0.05$) and $\sigma (R^{-1})$ error ($\pm 0.3$ TV$^{-1}$).

\subsubsection{Atmospheric Loss and Production Corrections} 

In order to obtain the absolute flux of primary protons and helium nuclei at TOA, corrections were applied for the survival probability 
in the residual atmosphere ($\eta$) and 
estimated atmospheric secondary production ratio ($R_{\rm air}$).
Both were estimated
by solving simultaneous transport equations \citep{Papini1996}.
The primary spectrum at TOA ($\Phi_{\rm TOA}$) was determined in an iterative procedure 
so that the estimated spectrum at TOI ($\Phi_{\rm TOI}$) agreed with the observed spectrum.
For BESS-Polar I, with residual atmosphere of 4.33 g/cm$^2$, the survival probability at 10 GeV/nucleon was estimated as 
$(94.42 \pm 0.56) \%$ for protons and
$(90.77 \pm 1.2) \%$ for helium. For BESS-Polar II, with residual atmosphere of 5.81 g/cm$^2$, the survival probability was estimated as 
$(92.47 \pm 0.74) \%$ for protons and
$(87.63 \pm 1.5) \%$ for helium. 
The proton secondary-to-primary ratio at 10 GeV was calculated as
$0.0174 \pm 0.0020$ in BESS-Polar I and
$0.0227 \pm 0.0026$ in BESS-Polar II.
Atmospheric secondary helium nuclei above 1 GeV/nucleon are dominated by the fragments of heavier
cosmic-ray nuclei. The secondary-to-primary helium ratio at 10 GeV/nucleon was calculated as 
$0.0209 \pm 0.0043$ in BESS-Polar I and 
$0.0271 \pm 0.0054$ in BESS-Polar II.
The errors in atmospheric correction were estimated by multiplying the factors which represent the uncertainty of residual air depth as $10 \%$, cross section of primary protons as $5 \%$, and the cross section of primary heavier cosmic ray nuclei as $20 \%$. 

\subsubsection{Top of Atmosphere Flux} 

Utilizing these atmospheric loss and production calculations, the  measured differential flux of a given particle species extrapolated at the top of the atmosphere ($\Phi_{\rm TOA}$) and binned in energy with width d$E$ is expressed as: 

\begin{equation}
\Phi_{\rm TOA} (E_{\rm TOA}) = \frac{\Phi_{\rm TOI} (E_{\rm TOI})}{\eta(E_{ATM})+R_{\rm air}(E_{ATM})}\\
\end{equation}

\noindent
where $E_{ATM}$ is a shorthand for the energy used in each step of the iterative process described above in which the energy of the particle passing through the atmosphere is calculated at 0.1 $g/cm^2$ intervals. The loss and production calculations are made over the same intervals using the corresponding $E_{ATM}$.

\section{Systematic Uncertainties}

\begin{figure}[!t]
 \centering
 \includegraphics[width=0.5\textwidth]{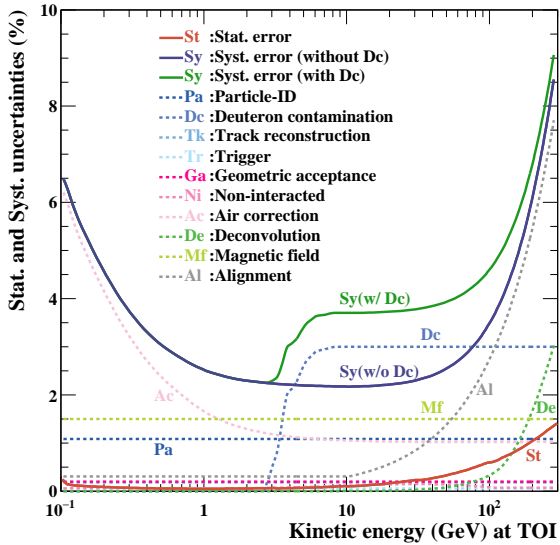}
 \caption{Individual and total systematic uncertainties and the statistical uncertainty for BESS-Polar II proton measurements. The total systematic uncertainty for protons is asymmetric. The lower (subtractive) uncertainty includes deuteron contamination but the upper (additive) does not.}
\label{Fig:syst}
\end{figure}

The overall systematic uncertainties for protons and helium share nine components; particle identification, track reconstruction, trigger, geometric acceptance, non-interacted single track selection, atmospheric production and loss correction, spectral deconvolution, magnetic field details, and tracking system alignment. For protons, deuteron contamination may also contribute by increasing the measured proton flux. Details of those factors are discussed above and their relative values as functions of particle energy are illustrated in
Figure \ref{Fig:syst} which shows the systematic and statistical uncertainties in BESS-Polar II proton measurements. The total systematic uncertainty in the proton measurements is asymmetric and is shown both with and without deuteron contamination. The upper (additive) total systematic uncertainty in the measured proton spectrum does not include deuteron contamination. The lower (subtractive) total systematic uncertainty in the measured proton spectrum includes deuteron contamination. The proton total systematic uncertainties at 10 GeV are
$1.9\%$ (upper), $3.4\%$ (lower) in BESS-Polar I and
$2.1\%$ (upper), $3.7\%$ (lower) in BESS-Polar II. The difference arises from the greater atmospheric overburden in BESS-Polar II.
The helium total systematic uncertainties are symmetric above and below the measured spectral points. At 10 GeV/nucleon they are 
$3.0\%$ in BESS-Polar I, and  
$3.2\%$ in BESS-Polar II.
However, the uncertainty of the non-interacted single track selection for helium is worse at high energy in BESS-Polar I because it is difficult to simulate the effect of $\delta$-rays and interactions for TOF paddles with only one PMT. Thus, at 80 GeV/nucleon the helium total systematic uncertainties are 
$5.3\%$ in BESS-Polar I, and  
$5.0\%$ in BESS-Polar II.

\section{Measured Spectra}

\begin{figure*}[!t]
 \centering
 \includegraphics[width=1.0\textwidth]{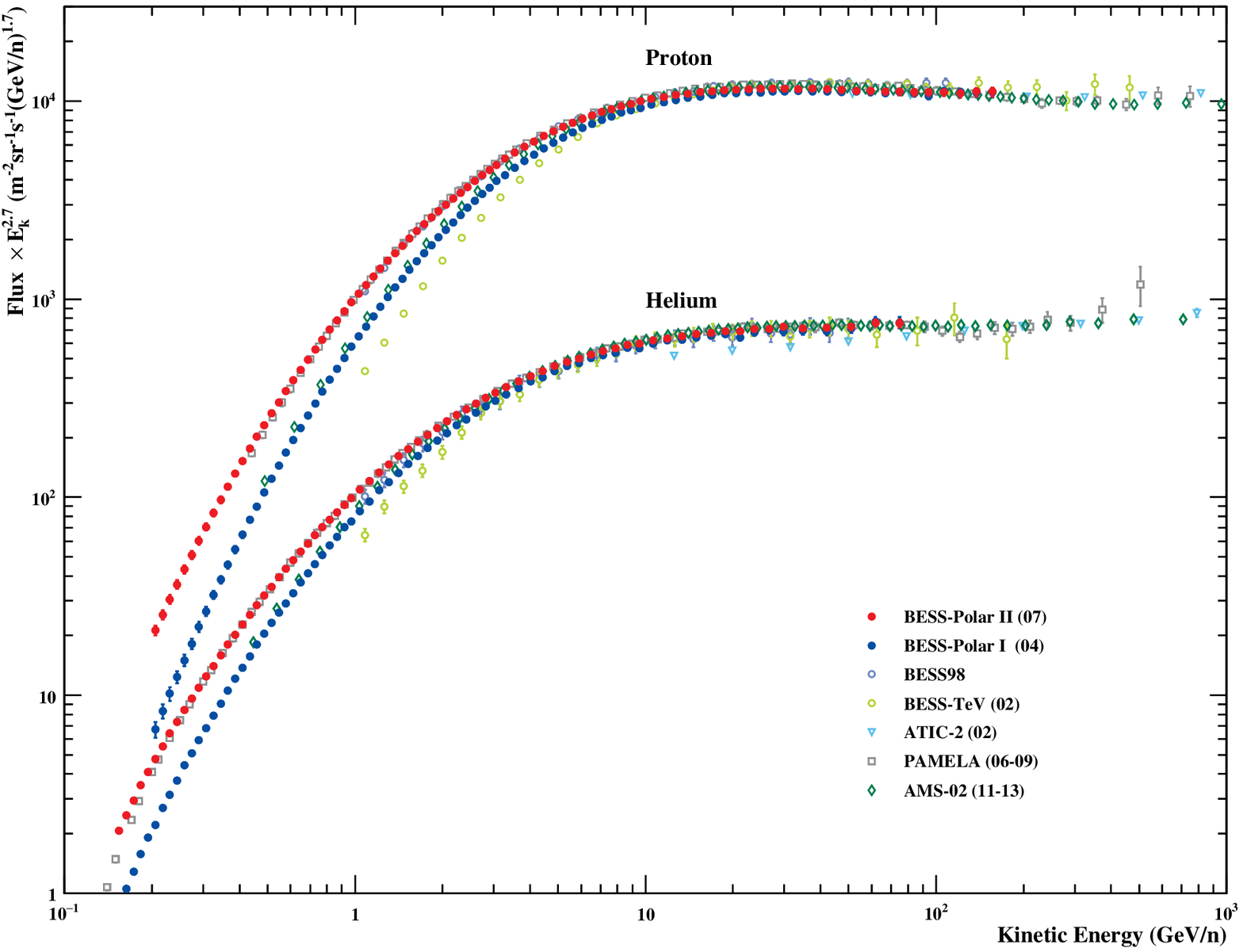}
 \caption{Absolute differential energy spectra of primary protons and helium nuclei at TOA multiplied by $E_{K}^{2.7}$ measured by BESS-Polar I and BESS-Polar II. Spectra reported by BESS98 \citep{0004-637X-545-2-1135}, BESS-TeV \citep{Haino200435}, ATIC-2 \citep{Panov2007}, PAMELA \citep{Adriani01042011,Adriani2014323}, and AMS-02 \citep{PhysRevLett.114.171103,PhysRevLett.115.211101} are shown for comparison. Differences in the BESS-Polar I and BESS-Polar II spectra at low energies are due to solar modulation.}
 \label{Fig:phe_7_27All}
\end{figure*}

We have measured the absolute TOA fluxes of protons 0.2-160 GeV
and helium nuclei 0.15-80 GeV/nucleon
using both BESS-Polar I and BESS-Polar II, where the lower energy was set by the requirement for particles to penetrate to the LTOF and upper energy was chosen to give flat detection efficiency. The corresponding ranges are 0.6-160 GV for protons and 1.1-160 GV for helium. The BESS-Polar measurements are summarized in Tables \ref{Table:proton} and
\ref{Table:helium}. 
The overall uncertainties, including both statistical and systematic errors, are less than
$\pm 10\%$ for BESS-Polar I protons, 
$\pm 6\%$ for BESS-Polar II protons, 
$\pm 7\%$ for BESS-Polar I helium, and 
$\pm 5\%$ for BESS-Polar II helium. 
The TOA proton and helium spectra are shown in Figure \ref{Fig:phe_7_27All}
in comparison with BESS98 \citep{0004-637X-545-2-1135}, BESS-TeV \citep{Haino200435}, ATIC-2 \citep{Panov2007}, PAMELA \citep{Adriani01042011,Adriani2014323}, and AMS-02 \citep{PhysRevLett.114.171103,PhysRevLett.115.211101}. We note that the BESS98 and BESS-Polar flights took place at very low and nearly constant geomagnetic cutoff. Their measured spectra do not have the systematic uncertainties at lower energies that result from the rigidity-dependent live time required by orbiting instruments such as PAMELA (70$^{\circ}$ inclination), and AMS-02 (51.6$^{\circ}$ inclination) to account for measurements in varying geomagnetic cutoff. CREAM proton and helium spectra \citep{2041-8205-714-1-L89} are not shown in Figure \ref{Fig:phe_7_27All} because most points are at higher energies.

\subsection{High Energy Interstellar Spectra} 
At energies above  30 GeV for protons and 15 GeV/nucleon for helium, the effects of solar modulation (discussed in Section \ref{ssec:le}) are effectively negligible and the proton spectra measured BESS-Polar I and BESS-Polar II are essentially identical, differing by less than $1\%$ at 160 GeV despite having been obtained during quite different periods of solar activity (see further discussion below). The same is true of the helium spectra above $\sim$15 GeV/nucleon where the BESS-Polar I and BESS-Polar II results differ by less than $1\%$ at 80 GeV/nucleon.  Thus, the interstellar spectra can be assumed to be the same as the measured TOA spectra over 30-160 GeV for protons and 15-80 GeV/nucleon for helium. Over these ranges, the measured spectra appear to follow single power laws. Thus, a given spectral $F$ can be parameterized by a power law in kinetic energy,
$E_{k}$, as $F(E_{k}) = \Phi_{E_{k}} E_{k}^{-\gamma^{E_{k}}}$ 
where $\Phi_{E_{k}}$ is the normalization constant.
For the results reported here, systematic errors related to spectral index (deconvolution and alignment) were taken into account, and a fitting range of 30-160 GeV was used
for protons and 15-80 GeV/nucleon for helium. The best fit values and uncertainties for protons ($\gamma_{\rm p}^{E_{k}}$)
and helium nuclei ($\gamma_{\rm He}^{E_{k}}$) for the two flights are:

\begin{fleqn}
\begin{align*}
\gamma_{{\rm p (Polar I)}}^{E_{k}} =& -2.715 \pm 0.006({\rm stat}) \pm 0.010({\rm syst})  \\
\gamma_{{\rm p (Polar II)}}^{E_{k}} =& -2.736 \pm 0.002({\rm stat}) \pm 0.010({\rm syst}) \\
\gamma_{{\rm He (Polar I)}}^{E_{k}} =& -2.606 \pm 0.017({\rm stat}) \pm 0.012({\rm syst}) \\
\gamma_{{\rm He (Polar II)}}^{E_{k}} =& -2.627 \pm 0.006({\rm stat}) \pm 0.012({\rm syst})
\end{align*}
\end{fleqn}

For both protons and helium, fits can be made over 30-160 GV to a single power law in $R$, parameterized as $F(R) = \Phi_{R} R^{-\gamma^{R}}$.  
The best fit values and uncertainties over a fitting range of 30-160 GV for both protons ($\gamma_{\rm p}^{R}$)
helium ($\gamma_{\rm He}^{R}$) for the two flights are: 

\begin{fleqn}
\begin{align*}
\gamma_{{\rm p (Polar I)}}^{R} =& -2.763 \pm 0.006({\rm stat}) \pm 0.010({\rm syst})  \\
\gamma_{{\rm p (Polar II)}}^{R} =& -2.784 \pm 0.002({\rm stat}) \pm 0.010({\rm syst}) \\
\gamma_{{\rm He (Polar I)}}^{R} =& -2.690 \pm 0.016({\rm stat}) \pm 0.011({\rm syst}) \\
\gamma_{{\rm He (Polar II)}}^{R} =& -2.709 \pm 0.006({\rm stat}) \pm 0.011({\rm syst})
\end{align*}
\end{fleqn}

At high-energy, PAMELA \citep{Adriani01042011,Adriani2014323}, AMS-02 \citep{PhysRevLett.114.171103,PhysRevLett.115.211101} and BESS-Polar proton and helium spectra agree within one $\sigma_{E}$. We note, however, that the BESS-Polar spectra do not extend to high enough energy to test the spectral hardening reported by PAMELA, ATIC-2 \citep{Panov2007}, CREAM \citep{2041-8205-714-1-L89} and AMS-02.

\begin{figure}[!t]
 \centering
 \includegraphics[width=0.50\textwidth]{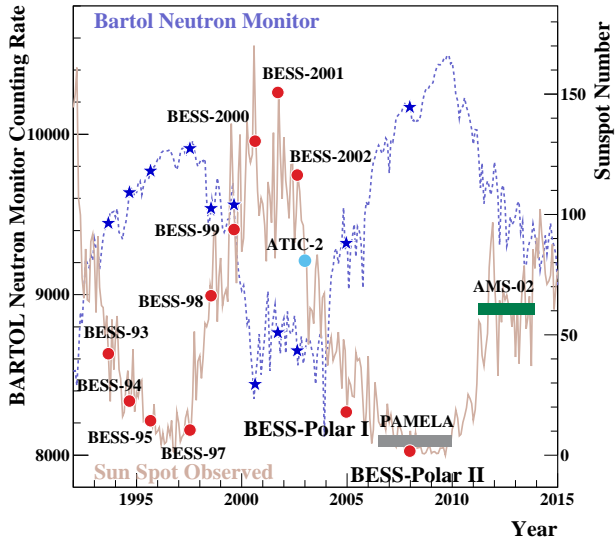}
 \caption{Variation of Bartol neutron monitor counts and sunspot number from 1991 to 2015. The BESS, ATIC-2 and BESS-Polar flights are marked as circles at the corresponding sunspot number. The periods for which PAMELA and AMS-02 proton and helium spectra have been reported are shown as bars at the approximate average sunspot number. The BESS-Polar II flight was carried out very near the deepest solar minimum.}
 \label{Fig:Solar}
\end{figure}

\subsection{Low Energies and Solar Modulation} \label{ssec:le}
GCR entering the heliosphere are scattered by
irregularities in the heliospheric magnetic field (HMF) and undergo
convection and adiabatic deceleration in the expanding solar wind.
The resulting modification of the GCR energy spectra, is known as
``solar modulation''.
Reflecting variations in solar activity and HMF polarity, large differences at low energies have been observed in the proton and 
helium spectra measured by the BESS collaboration since 1993. The solar cycle includes both a sunspot activity period of $\sim$11 years and a $\sim$22 year magnetic cycle with alternating positive ($A{>}0$) and
negative ($A{<}0$) phases. $A{<}0$ polarity cycles are defined as
the periods when the HMF is directed
towards the Sun in its northern hemisphere. HMF polarity reverses at a time when solar activity is maximum,
and the global magnetic field profile also reverses throughout the heliosphere with time lag due to propagation of the fields.
Positive and negative particles traversing the HMF drift in opposite directions,
taking different routes to arrive at the Earth and giving rise to ``flat'' and ``peaked'' periods in neutron
monitor data around ``solar minimum''. Thus, the details of the effect of the solar wind 
and its entrained magnetic fields on the incoming GCR have to be taken 
into account in deriving interstellar spectra at lower energies. At energies below 30 GeV for protons and 15 GeV/nucleon for helium (discussed above) we report only the measured TOA fluxes. However, we note the rich information on solar modulation that can be derived from BESS and BESS-Polar antiproton, proton, helium, and light isotope spectra obtained over more than a full solar cycle, including a reversal of solar magnetic polarity.
To illustrate the range of solar activity spanned by BESS and BESS-Polar, Figure~\ref{Fig:Solar} shows the Bartol neutron monitor
counting rate (blue dashed curve) \citep{Bieber2014} and the number of sunspots (orange curve) \citep{Hathaway2015} together the data of BESS flights (red circles and blue stars). The ATIC-2 (light blue circle), PAMELA (gray bar) and AMS-02 (green bar) flights are also indicated.

\begin{figure}[!t]
 \centering
 \includegraphics[width=0.5\textwidth]{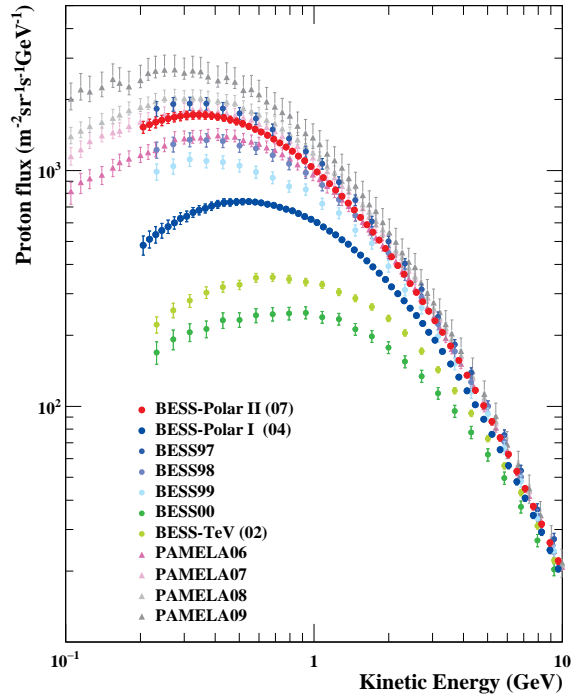}
 \caption{Absolute differential energy spectra of primary protons measured by BESS-Polar I and BESS-Polar II together with earlier BESS measurements and the PAMELA spectra compiled yearly  \citep{0004-637X-765-2-91}. The effect of solar modulation is evident at lower energies. Low energy measurements at similar solar modulation, e.g. BESS97, BESS-Polar II, and PAMELA07 are consistent within reported errors. The AMS-02 results in Figure \ref{Fig:phe_7_27All} are not reported by time period. }
 \label{Fig:p7}
\end{figure}

\begin{figure}[!t]
 \centering
 \includegraphics[width=0.5\textwidth]{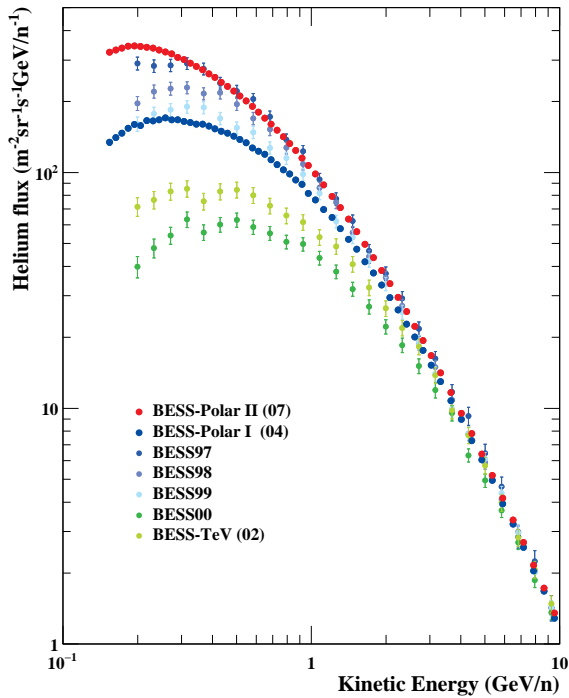}
 \caption{Absolute differential energy spectra of helium nuclei measured by BESS-Polar I and BESS-Polar II together with earlier BESS measurements. BESS97 and BESS-Polar II measurements at similar solar modulation are consistent within reported errors except at the lowest energies.} 

 \label{Fig:he7}
\end{figure}

Differences in the spectra measured by BESS-Polar I and BESS-Polar II
from solar activity below $\sim$10 GeV for protons in Figure \ref{Fig:p7} and below $\sim$5 GeV/nucleon for helium in Figure \ref{Fig:he7} are highlighted and compared to measurements by earlier BESS instruments and to PAMELA measurements for four years before and during the most recent solar minimum.
Varying solar modulation strongly affected proton and helium spectra measured by BESS (BESS 97, 98, 99) approaching the HMF polarity reversal from positive to negative in 2000, prior to the BESS 00 flight. Proton and helium measurements following the HMF polarity reversal (BESS 00, BESS-TeV, BESS-Polar I, BESS-Polar II) show much smaller variations with solar activity. This is consistent with the effect expected from charge-sign dependent drift \citep{PhysRevLett.83.674, PhysRevLett.88.051101, Potgieter20141415}. The proton spectra observed by BESS at solar minimum in 1997 and and by BESS-Polar II and PAMELA in 2007, near the next solar minimum are very similar. As would be expected, the low-energy PAMELA proton spectrum from 2006, prior to solar minimum falls below BESS-Polar II or PAMELA in 2007, and the PAMELA results from 2008 and 2009 during the deepest solar minimum are higher. The BESS and BESS-Polar measurements of antiprotons and protons, which differ only in charge-sign and interstellar spectral shape, provide an excellent test case for charge-sign dependent solar modulation \citep{PhysRevLett.88.051101}. A full discussion solar modulation and of the implications of the BESS results is beyond the scope of the present paper.

\subsection{Proton-to-Helium Ratio}

\begin{figure}[!t]
 \centering
 \includegraphics[width=0.5\textwidth]{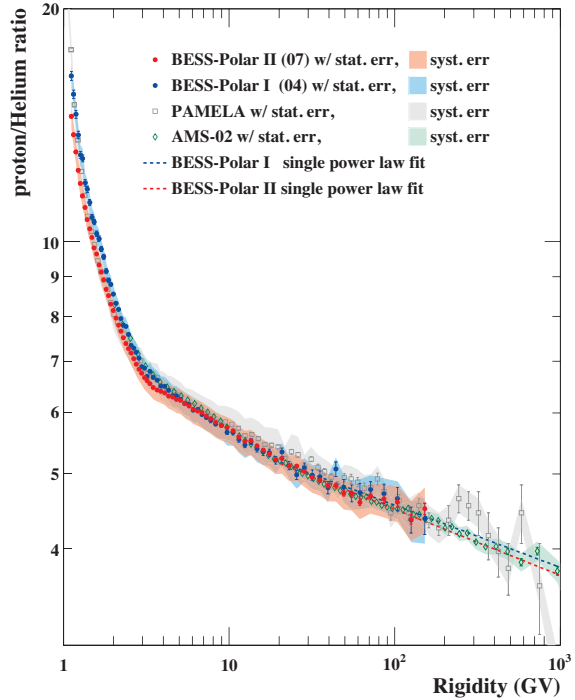}
 \caption{Proton/helium flux ratios measured by BESS-Polar I and BESS-Polar II (with protons rebinned to match helium). The ratio reported by PAMELA \citep{Adriani01042011,Adriani2014323} and the AMS-02 ratio \citep{PhysRevLett.115.211101} are also shown. Above 10 GV the two BESS-Polar ratios and AMS-02 agree well within error bars, while the PAMELA ratio is slightly higher and flatter with increasing $R$.}
 \label{Fig:ratio}
\end{figure}

Comparing the spectra of protons and helium gives important information on the sources and acceleration of the light cosmic-rays and a test of cosmic-ray origin and propagation models. The measured power-law indices of protons and helium, given above, are slightly different. This is most easily shown in $R$ space because our measurements for both species reach 160 GV. The measured proton/helium flux ratios in $R$ from BESS-Polar I and BESS-Polar II, with protons re-binned to match helium, are shown in Figure \ref{Fig:ratio} and compared to PAMELA \citep{Adriani01042011,Adriani2014323} data and with AMS-02 results \citep{PhysRevLett.115.211101}. The BESS-Polar proton/helium flux ratios are summarized in Table \ref{Table:ratio}. A simple model of solar modulation, appropriate when comparing two species of the same charge-sign, shows that the effect of solar modulation on this ratio should be negligible above 20 GV. This is borne out by the measured ratios. As expected, p/He above 20 GV is well represented by a single power-law with an index which is just the difference of the indices of the protons and helium as
($\Delta_{\gamma^{R}} = \gamma_{\rm p}^{R} - \gamma_{\rm He}^{R}$). This was also determined directly using a power-law fit to p/He above 20 GV, as shown in Figure \ref{Fig:ratio}.
The fit values obtained were
$\Delta_{\gamma^{R-fit}} = -0.081 \pm 0.010({\rm stat}) \pm 0.020({\rm syst})$
for BESS-Polar I and 
$\Delta_{\gamma^{R-fit}} = -0.086 \pm 0.004({\rm stat}) \pm 0.022({\rm syst})$
for BESS Polar II. These compare well to the calculated values of 
$\Delta_{\gamma^{R-calc}} = -0.073 \pm 0.017({\rm stat}) \pm 0.015({\rm syst})$
for BESS-Polar I and 
$\Delta_{\gamma^{R-calc}} = -0.075 \pm 0.006({\rm stat}) \pm 0.015({\rm syst})$
for BESS-Polar II. 
Above 10 GV the two BESS-Polar ratios and AMS-02 agree well within error bars, while the PAMELA ratio is slightly higher and flatter with increasing $R$.

\section{Conclusion}

With the BESS-Polar I long-duration balloon flight in 2004 and the BESS-Polar II flight in 2007/2008, we  have measured the energy spectra at the top of the atmosphere of protons in the range 
0.2-160 GeV and helium nuclei in the range 0.15-80 GeV/nucleon, corresponding to magnetic-rigidity ranges of 0.6-160 GV for protons and 1.1-160 GV for helium. At energies above the effects of solar modulation, the spectra of either species measured in the two flights differ by less than $1\%$ at the highest measured energy (rigidity). Within the measured ranges, the high energy (rigidity) spectra of both species can be represented very well by single power laws. We have also determined the measured proton to helium ratio and show that at high energies it is also fit by a single power law.

\section*{Acknowlegments}
 The BESS-Polar program is a Japan-United States collaboration, supported in Japan
by the Grant-in-Aid ``KAKENHI" for Specially Promoted and Basic Researches, MEXT-JSPS, 
and in the United States by NASA. 
Balloon flight operations were carried out by the NASA Columbia Scientific Balloon (CSBF) 
Facility and the National Science Foundation United States Antarctic Program (USAP). 
We express our sincere thanks for the financial support and encouragement of both national agencies and the continuous and professional support of the technical and administrative staffs of the collaborating institutions and of CSBF and NSF USAP.

\bibliography{ref}

\begin{deluxetable}{llll}
\tabletypesize{\scriptsize}
\tablewidth{0pc}
\tablecolumns{4}
\tablecaption{Proton flux at the top of atmosphere}
\tablehead{
\colhead{} & \colhead{} & 
\colhead{ BESS-Polar I} & 
\colhead{ BESS-Polar II} \\
\cline{3-3} \cline{4-4} \vspace{-0.2cm} \\
\colhead{Energy Range} \vspace{0.05cm} & \colhead{ $E_{k}$ } & 
\colhead{${\rm Flux} \pm \Delta F_{ \rm sta} \pm \Delta F_{\rm sys}$ } &  
\colhead{${\rm Flux} \pm \Delta F_{ \rm sta} \pm \Delta F_{\rm sys}$ } \\ 
\colhead{ (GeV) } & \colhead{ (GeV) } & 
\colhead{(m$^{2}$ sr s GeV)$^{-1}$ } &
\colhead{(m$^{2}$ sr s GeV)$^{-1}$ }
}
\input{table01}
\label{Table:proton}
\end{deluxetable}

\begin{deluxetable}{llll}
\tabletypesize{\scriptsize}
\tablewidth{0pt}
\tablecolumns{4}
\tablecaption{Helium flux at the top of atmosphere}
\tablehead{
\colhead{} & \colhead{} & 
\colhead{ BESS-Polar I} & 
\colhead{ BESS-Polar II} \\
\cline{3-3} \cline{4-4} \vspace{-0.2cm} \\
\colhead{Energy Range} \vspace{0.05cm} & \colhead{ $E_{k}$ } & 
\colhead{${\rm Flux} \pm \Delta F_{ \rm sta} \pm \Delta F_{\rm sys}$ } &  
\colhead{${\rm Flux} \pm \Delta F_{ \rm sta} \pm \Delta F_{\rm sys}$ } \\ 
\colhead{ (GeV/nucleon) } & \colhead{ (GeV/nucleon) } & 
\colhead{(m$^{2}$ sr s GeV/nucleon)$^{-1}$ } & 
\colhead{(m$^{2}$ sr s GeV/nucleon)$^{-1}$ }
}
\input{table02}
\label{Table:helium}
\end{deluxetable}

\begin{deluxetable}{llll}
\tabletypesize{\scriptsize}
\tablewidth{0pt}
\tablecolumns{4}
\tablecaption{Proton/Helium ratio at the top of atmosphere}
\tablehead{
\colhead{} & \colhead{} & 
\colhead{ BESS-Polar I} & 
\colhead{ BESS-Polar II} \\
\cline{3-3} \cline{4-4} \vspace{-0.2cm} \\
\colhead{Rigidity Range} \vspace{0.05cm} & \colhead{ $R$ } & 
\colhead{${\rm Ratio} \pm \Delta RO_{ \rm sta} \pm \Delta RO_{\rm sys}$ } &  
\colhead{${\rm Ratio} \pm \Delta RO_{ \rm sta} \pm \Delta RO_{\rm sys}$ } \\ 
\colhead{ (GV) } & \colhead{ (GV) } & 
\colhead{(one)} & 
\colhead{(one)}
}
\input{table03}
\label{Table:ratio}
\end{deluxetable}

\end{document}

%% file: table01.tex
\startdata
0.200--0.211 & 0.205 & $4.825 \pm 0.032 \pm 0.446 \times 10^{2}$ & $1.527 \pm 0.002 \pm 0.089 \times 10^{3}$ \\
0.211--0.224 & 0.218 & $5.110 \pm 0.031 \pm 0.422 \times 10^{2}$ & $1.567 \pm 0.002 \pm 0.086 \times 10^{3}$ \\
0.224--0.237 & 0.231 & $5.353 \pm 0.030 \pm 0.423 \times 10^{2}$ & $1.603 \pm 0.002 \pm 0.086 \times 10^{3}$ \\
0.237--0.251 & 0.244 & $5.565 \pm 0.029 \pm 0.397 \times 10^{2}$ & $1.633 \pm 0.002 \pm 0.082 \times 10^{3}$ \\
0.251--0.266 & 0.259 & $5.780 \pm 0.028 \pm 0.374 \times 10^{2}$ & $1.662 \pm 0.002 \pm 0.079 \times 10^{3}$ \\
0.266--0.282 & 0.274 & $6.000 \pm 0.027 \pm 0.354 \times 10^{2}$ & $1.684 \pm 0.001 \pm 0.076 \times 10^{3}$ \\
0.282--0.298 & 0.290 & $6.263 \pm 0.026 \pm 0.358 \times 10^{2}$ & $1.702 \pm 0.001 \pm 0.075 \times 10^{3}$ \\
0.298--0.316 & 0.307 & $6.399 \pm 0.025 \pm 0.336 \times 10^{2}$ & $1.713 \pm 0.001 \pm 0.072 \times 10^{3}$ \\
0.316--0.335 & 0.326 & $6.649 \pm 0.024 \pm 0.322 \times 10^{2}$ & $1.722 \pm 0.001 \pm 0.069 \times 10^{3}$ \\
0.335--0.355 & 0.345 & $6.787 \pm 0.024 \pm 0.304 \times 10^{2}$ & $1.723 \pm 0.001 \pm 0.066 \times 10^{3}$ \\
0.355--0.376 & 0.365 & $6.910 \pm 0.023 \pm 0.303 \times 10^{2}$ & $1.718 \pm 0.001 \pm 0.065 \times 10^{3}$ \\
0.376--0.398 & 0.387 & $7.061 \pm 0.022 \pm 0.288 \times 10^{2}$ & $1.708 \pm 0.001 \pm 0.062 \times 10^{3}$ \\
0.398--0.422 & 0.410 & $7.191 \pm 0.021 \pm 0.275 \times 10^{2}$ & $1.693 \pm 0.001 \pm 0.059 \times 10^{3}$ \\
0.422--0.447 & 0.434 & $7.319 \pm 0.021 \pm 0.262 \times 10^{2}$ & $1.673 \pm 0.001 \pm 0.056 \times 10^{3}$ \\
0.447--0.473 & 0.460 & $7.326 \pm 0.020 \pm 0.259 \times 10^{2}$ & $1.645 \pm 0.001 \pm 0.055 \times 10^{3}$ \\
0.473--0.501 & 0.487 & $7.373 \pm 0.019 \pm 0.246 \times 10^{2}$ & $1.615 \pm 0.001 \pm 0.052 \times 10^{3}$ \\
0.501--0.531 & 0.516 & $7.386 \pm 0.019 \pm 0.234 \times 10^{2}$ & $1.582 \pm 0.001 \pm 0.050 \times 10^{3}$ \\
0.531--0.562 & 0.546 & $7.391 \pm 0.018 \pm 0.222 \times 10^{2}$ & $1.542 \pm 0.001 \pm 0.047 \times 10^{3}$ \\
0.562--0.596 & 0.579 & $7.351 \pm 0.017 \pm 0.219 \times 10^{2}$ & $1.501 \pm 0.001 \pm 0.045 \times 10^{3}$ \\
0.596--0.631 & 0.613 & $7.295 \pm 0.017 \pm 0.208 \times 10^{2}$ & $1.456 \pm 0.001 \pm 0.043 \times 10^{3}$ \\
0.631--0.668 & 0.649 & $7.189 \pm 0.016 \pm 0.196 \times 10^{2}$ & $1.411 \pm 0.001 \pm 0.040 \times 10^{3}$ \\
0.668--0.708 & 0.688 & $7.116 \pm 0.015 \pm 0.187 \times 10^{2}$ & $1.360 \pm 0.001 \pm 0.038 \times 10^{3}$ \\
0.708--0.750 & 0.729 & $6.994 \pm 0.015 \pm 0.182 \times 10^{2}$ & $1.308 \pm 0.001 \pm 0.036 \times 10^{3}$ \\
0.750--0.794 & 0.772 & $6.862 \pm 0.014 \pm 0.172 \times 10^{2}$ & $1.256 \pm 0.001 \pm 0.034 \times 10^{3}$ \\
0.794--0.841 & 0.818 & $6.746 \pm 0.013 \pm 0.163 \times 10^{2}$ & $1.207 \pm 0.001 \pm 0.032 \times 10^{3}$ \\
0.841--0.891 & 0.866 & $6.571 \pm 0.013 \pm 0.154 \times 10^{2}$ & $1.153 \pm 0.001 \pm 0.029 \times 10^{3}$ \\
0.891--0.944 & 0.917 & $6.388 \pm 0.012 \pm 0.149 \times 10^{2}$ & $1.099 \pm 0.001 \pm 0.028 \times 10^{3}$ \\
0.944--1.000 & 0.972 & $6.218 \pm 0.012 \pm 0.142 \times 10^{2}$ & $1.041 \pm 0.001 \pm 0.026 \times 10^{3}$ \\
1.000--1.059 & 1.029 & $6.011 \pm 0.011 \pm 0.134 \times 10^{2}$ & $9.875 \pm 0.005 \pm 0.243 \times 10^{2}$ \\
1.059--1.122 & 1.090 & $5.760 \pm 0.011 \pm 0.125 \times 10^{2}$ & $9.327 \pm 0.005 \pm 0.225 \times 10^{2}$ \\
1.122--1.189 & 1.155 & $5.535 \pm 0.010 \pm 0.120 \times 10^{2}$ & $8.798 \pm 0.005 \pm 0.212 \times 10^{2}$ \\
1.189--1.259 & 1.223 & $5.336 \pm 0.010 \pm 0.114 \times 10^{2}$ & $8.276 \pm 0.004 \pm 0.197 \times 10^{2}$ \\
1.259--1.334 & 1.296 & $5.078 \pm 0.009 \pm 0.107 \times 10^{2}$ & $7.766 \pm 0.004 \pm 0.182 \times 10^{2}$ \\
1.334--1.413 & 1.373 & $4.869 \pm 0.009 \pm 0.101 \times 10^{2}$ & $7.270 \pm 0.004 \pm 0.168 \times 10^{2}$ \\
1.413--1.496 & 1.454 & $4.607 \pm 0.008 \pm 0.095 \times 10^{2}$ & $6.794 \pm 0.004 \pm 0.157 \times 10^{2}$ \\
1.496--1.585 & 1.540 & $4.378 \pm 0.008 \pm 0.089 \times 10^{2}$ & $6.333 \pm 0.003 \pm 0.145 \times 10^{2}$ \\
1.585--1.679 & 1.631 & $4.143 \pm 0.007 \pm 0.083 \times 10^{2}$ & $5.896 \pm 0.003 \pm 0.133 \times 10^{2}$ \\
1.679--1.778 & 1.728 & $3.905 \pm 0.007 \pm 0.078 \times 10^{2}$ & $5.473 \pm 0.003 \pm 0.123 \times 10^{2}$ \\
1.778--1.884 & 1.830 & $3.667 \pm 0.006 \pm 0.073 \times 10^{2}$ & $5.065 \pm 0.003 \pm 0.113 \times 10^{2}$ \\
1.884--1.995 & 1.939 & $3.447 \pm 0.006 \pm 0.068 \times 10^{2}$ & $4.673 \pm 0.003 \pm 0.104 \times 10^{2}$ \\
1.995--2.113 & 2.054 & $3.214 \pm 0.006 \pm 0.063 \times 10^{2}$ & $4.307 \pm 0.002 \pm 0.095 \times 10^{2}$ \\
2.113--2.239 & 2.175 & $3.003 \pm 0.005 \pm 0.058 \times 10^{2}$ & $3.956 \pm 0.002 \pm 0.086 \times 10^{2}$ \\
2.239--2.371 & 2.304 & $2.804 \pm 0.005 \pm 0.054 \times 10^{2}$ & $3.628 \pm 0.002 \pm 0.079 \times 10^{2}$ \\
2.371--2.512 & 2.441 & $2.607 \pm 0.005 \pm 0.050 \times 10^{2}$ & $3.318 \pm 0.002 \pm 0.072 \times 10^{2}$ \\
2.512--2.661 & 2.585 & $2.424 \pm 0.004 \pm 0.046 \times 10^{2}$ & $3.041 \pm 0.002 \pm 0.066 \times 10^{2}$ \\
2.661--2.818 & 2.739 & $2.242 \pm 0.004 ^{\raisebox{-0.2ex}[0ex][-1ex]{\hspace{2.5mm}\tiny{$+$0.043}}}_{\raisebox{0.1ex}{\hspace{2.4mm}\tiny{$-$0.043}}} \times 10^{2}$ & $2.777 \pm 0.002 ^{\raisebox{-0.2ex}[0ex][-1ex]{\hspace{2.5mm}\tiny{$+$0.060}}}_{\raisebox{0.1ex}{\hspace{2.4mm}\tiny{$-$0.060}}} \times 10^{2}$ \\
2.818--2.985 & 2.901 & $2.059 \pm 0.004 ^{\raisebox{-0.2ex}[0ex][-1ex]{\hspace{2.5mm}\tiny{$+$0.039}}}_{\raisebox{0.1ex}{\hspace{2.4mm}\tiny{$-$0.039}}} \times 10^{2}$ & $2.531 \pm 0.002 ^{\raisebox{-0.2ex}[0ex][-1ex]{\hspace{2.5mm}\tiny{$+$0.054}}}_{\raisebox{0.1ex}{\hspace{2.4mm}\tiny{$-$0.054}}} \times 10^{2}$ \\
2.985--3.162 & 3.073 & $1.906 \pm 0.004 ^{\raisebox{-0.2ex}[0ex][-1ex]{\hspace{2.5mm}\tiny{$+$0.036}}}_{\raisebox{0.1ex}{\hspace{2.4mm}\tiny{$-$0.037}}} \times 10^{2}$ & $2.305 \pm 0.001 ^{\raisebox{-0.2ex}[0ex][-1ex]{\hspace{2.5mm}\tiny{$+$0.049}}}_{\raisebox{0.1ex}{\hspace{2.4mm}\tiny{$-$0.051}}} \times 10^{2}$ \\
3.162--3.415 & 3.286 & $1.709 \pm 0.003 ^{\raisebox{-0.2ex}[0ex][-1ex]{\hspace{2.5mm}\tiny{$+$0.032}}}_{\raisebox{0.1ex}{\hspace{2.4mm}\tiny{$-$0.034}}} \times 10^{2}$ & $2.061 \pm 0.001 ^{\raisebox{-0.2ex}[0ex][-1ex]{\hspace{2.5mm}\tiny{$+$0.044}}}_{\raisebox{0.1ex}{\hspace{2.4mm}\tiny{$-$0.047}}} \times 10^{2}$ \\
3.415--3.687 & 3.548 & $1.513 \pm 0.003 ^{\raisebox{-0.2ex}[0ex][-1ex]{\hspace{2.5mm}\tiny{$+$0.028}}}_{\raisebox{0.1ex}{\hspace{2.4mm}\tiny{$-$0.033}}} \times 10^{2}$ & $1.801 \pm 0.001 ^{\raisebox{-0.2ex}[0ex][-1ex]{\hspace{2.5mm}\tiny{$+$0.038}}}_{\raisebox{0.1ex}{\hspace{2.4mm}\tiny{$-$0.046}}} \times 10^{2}$ \\
3.687--3.981 & 3.831 & $1.330 \pm 0.002 ^{\raisebox{-0.2ex}[0ex][-1ex]{\hspace{2.5mm}\tiny{$+$0.025}}}_{\raisebox{0.1ex}{\hspace{2.4mm}\tiny{$-$0.032}}} \times 10^{2}$ & $1.566 \pm 0.001 ^{\raisebox{-0.2ex}[0ex][-1ex]{\hspace{2.5mm}\tiny{$+$0.033}}}_{\raisebox{0.1ex}{\hspace{2.4mm}\tiny{$-$0.045}}} \times 10^{2}$ \\
3.981--4.299 & 4.137 & $1.164 \pm 0.002 ^{\raisebox{-0.2ex}[0ex][-1ex]{\hspace{2.5mm}\tiny{$+$0.022}}}_{\raisebox{0.1ex}{\hspace{2.4mm}\tiny{$-$0.029}}} \times 10^{2}$ & $1.356 \pm 0.001 ^{\raisebox{-0.2ex}[0ex][-1ex]{\hspace{2.5mm}\tiny{$+$0.029}}}_{\raisebox{0.1ex}{\hspace{2.4mm}\tiny{$-$0.040}}} \times 10^{2}$ \\
4.299--4.642 & 4.467 & $1.015 \pm 0.002 ^{\raisebox{-0.2ex}[0ex][-1ex]{\hspace{2.5mm}\tiny{$+$0.019}}}_{\raisebox{0.1ex}{\hspace{2.4mm}\tiny{$-$0.026}}} \times 10^{2}$ & $1.171 \pm 0.001 ^{\raisebox{-0.2ex}[0ex][-1ex]{\hspace{2.5mm}\tiny{$+$0.025}}}_{\raisebox{0.1ex}{\hspace{2.4mm}\tiny{$-$0.035}}} \times 10^{2}$ \\
4.642--5.012 & 4.823 & $8.816 \pm 0.017 ^{\raisebox{-0.2ex}[0ex][-1ex]{\hspace{2.5mm}\tiny{$+$0.163}}}_{\raisebox{0.1ex}{\hspace{2.4mm}\tiny{$-$0.240}}} \times 10^{1}$ & $1.006 \pm 0.001 ^{\raisebox{-0.2ex}[0ex][-1ex]{\hspace{2.5mm}\tiny{$+$0.021}}}_{\raisebox{0.1ex}{\hspace{2.4mm}\tiny{$-$0.032}}} \times 10^{2}$ \\
5.012--5.412 & 5.208 & $7.612 \pm 0.015 ^{\raisebox{-0.2ex}[0ex][-1ex]{\hspace{2.5mm}\tiny{$+$0.141}}}_{\raisebox{0.1ex}{\hspace{2.4mm}\tiny{$-$0.221}}} \times 10^{1}$ & $8.621 \pm 0.006 ^{\raisebox{-0.2ex}[0ex][-1ex]{\hspace{2.5mm}\tiny{$+$0.179}}}_{\raisebox{0.1ex}{\hspace{2.4mm}\tiny{$-$0.285}}} \times 10^{1}$ \\
5.412--5.843 & 5.623 & $6.557 \pm 0.013 ^{\raisebox{-0.2ex}[0ex][-1ex]{\hspace{2.5mm}\tiny{$+$0.121}}}_{\raisebox{0.1ex}{\hspace{2.4mm}\tiny{$-$0.198}}} \times 10^{1}$ & $7.364 \pm 0.005 ^{\raisebox{-0.2ex}[0ex][-1ex]{\hspace{2.5mm}\tiny{$+$0.153}}}_{\raisebox{0.1ex}{\hspace{2.4mm}\tiny{$-$0.248}}} \times 10^{1}$ \\
5.843--6.310 & 6.072 & $5.608 \pm 0.012 ^{\raisebox{-0.2ex}[0ex][-1ex]{\hspace{2.5mm}\tiny{$+$0.103}}}_{\raisebox{0.1ex}{\hspace{2.4mm}\tiny{$-$0.176}}} \times 10^{1}$ & $6.257 \pm 0.005 ^{\raisebox{-0.2ex}[0ex][-1ex]{\hspace{2.5mm}\tiny{$+$0.130}}}_{\raisebox{0.1ex}{\hspace{2.4mm}\tiny{$-$0.215}}} \times 10^{1}$ \\
6.310--6.813 & 6.556 & $4.790 \pm 0.010 ^{\raisebox{-0.2ex}[0ex][-1ex]{\hspace{2.5mm}\tiny{$+$0.088}}}_{\raisebox{0.1ex}{\hspace{2.4mm}\tiny{$-$0.151}}} \times 10^{1}$ & $5.305 \pm 0.004 ^{\raisebox{-0.2ex}[0ex][-1ex]{\hspace{2.5mm}\tiny{$+$0.110}}}_{\raisebox{0.1ex}{\hspace{2.4mm}\tiny{$-$0.184}}} \times 10^{1}$ \\
6.813--7.356 & 7.079 & $4.083 \pm 0.009 ^{\raisebox{-0.2ex}[0ex][-1ex]{\hspace{2.5mm}\tiny{$+$0.075}}}_{\raisebox{0.1ex}{\hspace{2.4mm}\tiny{$-$0.130}}} \times 10^{1}$ & $4.481 \pm 0.004 ^{\raisebox{-0.2ex}[0ex][-1ex]{\hspace{2.5mm}\tiny{$+$0.093}}}_{\raisebox{0.1ex}{\hspace{2.4mm}\tiny{$-$0.156}}} \times 10^{1}$ \\
7.356--7.943 & 7.644 & $3.449 \pm 0.008 ^{\raisebox{-0.2ex}[0ex][-1ex]{\hspace{2.5mm}\tiny{$+$0.063}}}_{\raisebox{0.1ex}{\hspace{2.4mm}\tiny{$-$0.112}}} \times 10^{1}$ & $3.771 \pm 0.003 ^{\raisebox{-0.2ex}[0ex][-1ex]{\hspace{2.5mm}\tiny{$+$0.078}}}_{\raisebox{0.1ex}{\hspace{2.4mm}\tiny{$-$0.132}}} \times 10^{1}$ \\
7.943--8.577 & 8.254 & $2.930 \pm 0.007 ^{\raisebox{-0.2ex}[0ex][-1ex]{\hspace{2.5mm}\tiny{$+$0.054}}}_{\raisebox{0.1ex}{\hspace{2.4mm}\tiny{$-$0.096}}} \times 10^{1}$ & $3.169 \pm 0.003 ^{\raisebox{-0.2ex}[0ex][-1ex]{\hspace{2.5mm}\tiny{$+$0.065}}}_{\raisebox{0.1ex}{\hspace{2.4mm}\tiny{$-$0.111}}} \times 10^{1}$ \\
8.577--9.261 & 8.912 & $2.457 \pm 0.006 ^{\raisebox{-0.2ex}[0ex][-1ex]{\hspace{2.5mm}\tiny{$+$0.045}}}_{\raisebox{0.1ex}{\hspace{2.4mm}\tiny{$-$0.081}}} \times 10^{1}$ & $2.644 \pm 0.003 ^{\raisebox{-0.2ex}[0ex][-1ex]{\hspace{2.5mm}\tiny{$+$0.055}}}_{\raisebox{0.1ex}{\hspace{2.4mm}\tiny{$-$0.093}}} \times 10^{1}$ \\
9.261--10.00 & 9.623 & $2.050 \pm 0.006 ^{\raisebox{-0.2ex}[0ex][-1ex]{\hspace{2.5mm}\tiny{$+$0.037}}}_{\raisebox{0.1ex}{\hspace{2.4mm}\tiny{$-$0.067}}} \times 10^{1}$ & $2.211 \pm 0.002 ^{\raisebox{-0.2ex}[0ex][-1ex]{\hspace{2.5mm}\tiny{$+$0.045}}}_{\raisebox{0.1ex}{\hspace{2.4mm}\tiny{$-$0.077}}} \times 10^{1}$ \\
10.00--11.01 & 10.49 & $1.682 \pm 0.004 ^{\raisebox{-0.2ex}[0ex][-1ex]{\hspace{2.5mm}\tiny{$+$0.031}}}_{\raisebox{0.1ex}{\hspace{2.4mm}\tiny{$-$0.055}}} \times 10^{1}$ & $1.797 \pm 0.002 ^{\raisebox{-0.2ex}[0ex][-1ex]{\hspace{2.5mm}\tiny{$+$0.037}}}_{\raisebox{0.1ex}{\hspace{2.4mm}\tiny{$-$0.063}}} \times 10^{1}$ \\
11.01--12.12 & 11.55 & $1.336 \pm 0.004 ^{\raisebox{-0.2ex}[0ex][-1ex]{\hspace{2.5mm}\tiny{$+$0.024}}}_{\raisebox{0.1ex}{\hspace{2.4mm}\tiny{$-$0.044}}} \times 10^{1}$ & $1.419 \pm 0.001 ^{\raisebox{-0.2ex}[0ex][-1ex]{\hspace{2.5mm}\tiny{$+$0.029}}}_{\raisebox{0.1ex}{\hspace{2.4mm}\tiny{$-$0.050}}} \times 10^{1}$ \\
12.12--13.34 & 12.71 & $1.059 \pm 0.003 ^{\raisebox{-0.2ex}[0ex][-1ex]{\hspace{2.5mm}\tiny{$+$0.019}}}_{\raisebox{0.1ex}{\hspace{2.4mm}\tiny{$-$0.035}}} \times 10^{1}$ & $1.123 \pm 0.001 ^{\raisebox{-0.2ex}[0ex][-1ex]{\hspace{2.5mm}\tiny{$+$0.023}}}_{\raisebox{0.1ex}{\hspace{2.4mm}\tiny{$-$0.039}}} \times 10^{1}$ \\
13.34--14.68 & 13.99 & $8.393 \pm 0.027 ^{\raisebox{-0.2ex}[0ex][-1ex]{\hspace{2.5mm}\tiny{$+$0.154}}}_{\raisebox{0.1ex}{\hspace{2.4mm}\tiny{$-$0.277}}}$ & $8.802 \pm 0.010 ^{\raisebox{-0.2ex}[0ex][-1ex]{\hspace{2.5mm}\tiny{$+$0.182}}}_{\raisebox{0.1ex}{\hspace{2.4mm}\tiny{$-$0.309}}}$ \\
14.68--16.16 & 15.40 & $6.563 \pm 0.023 ^{\raisebox{-0.2ex}[0ex][-1ex]{\hspace{2.5mm}\tiny{$+$0.121}}}_{\raisebox{0.1ex}{\hspace{2.4mm}\tiny{$-$0.216}}}$ & $6.910 \pm 0.009 ^{\raisebox{-0.2ex}[0ex][-1ex]{\hspace{2.5mm}\tiny{$+$0.143}}}_{\raisebox{0.1ex}{\hspace{2.4mm}\tiny{$-$0.243}}}$ \\
16.16--17.78 & 16.95 & $5.156 \pm 0.019 ^{\raisebox{-0.2ex}[0ex][-1ex]{\hspace{2.5mm}\tiny{$+$0.095}}}_{\raisebox{0.1ex}{\hspace{2.4mm}\tiny{$-$0.170}}}$ & $5.393 \pm 0.007 ^{\raisebox{-0.2ex}[0ex][-1ex]{\hspace{2.5mm}\tiny{$+$0.112}}}_{\raisebox{0.1ex}{\hspace{2.4mm}\tiny{$-$0.190}}}$ \\
17.78--19.57 & 18.65 & $4.059 \pm 0.016 ^{\raisebox{-0.2ex}[0ex][-1ex]{\hspace{2.5mm}\tiny{$+$0.076}}}_{\raisebox{0.1ex}{\hspace{2.4mm}\tiny{$-$0.134}}}$ & $4.201 \pm 0.006 ^{\raisebox{-0.2ex}[0ex][-1ex]{\hspace{2.5mm}\tiny{$+$0.088}}}_{\raisebox{0.1ex}{\hspace{2.4mm}\tiny{$-$0.148}}}$ \\
19.57--21.54 & 20.53 & $3.147 \pm 0.014 ^{\raisebox{-0.2ex}[0ex][-1ex]{\hspace{2.5mm}\tiny{$+$0.059}}}_{\raisebox{0.1ex}{\hspace{2.4mm}\tiny{$-$0.104}}}$ & $3.271 \pm 0.005 ^{\raisebox{-0.2ex}[0ex][-1ex]{\hspace{2.5mm}\tiny{$+$0.069}}}_{\raisebox{0.1ex}{\hspace{2.4mm}\tiny{$-$0.116}}}$ \\
21.54--23.71 & 22.60 & $2.455 \pm 0.011 ^{\raisebox{-0.2ex}[0ex][-1ex]{\hspace{2.5mm}\tiny{$+$0.046}}}_{\raisebox{0.1ex}{\hspace{2.4mm}\tiny{$-$0.082}}}$ & $2.539 \pm 0.004 ^{\raisebox{-0.2ex}[0ex][-1ex]{\hspace{2.5mm}\tiny{$+$0.054}}}_{\raisebox{0.1ex}{\hspace{2.4mm}\tiny{$-$0.090}}}$ \\
23.71--26.10 & 24.87 & $1.883 \pm 0.010 ^{\raisebox{-0.2ex}[0ex][-1ex]{\hspace{2.5mm}\tiny{$+$0.036}}}_{\raisebox{0.1ex}{\hspace{2.4mm}\tiny{$-$0.063}}}$ & $1.972 \pm 0.004 ^{\raisebox{-0.2ex}[0ex][-1ex]{\hspace{2.5mm}\tiny{$+$0.042}}}_{\raisebox{0.1ex}{\hspace{2.4mm}\tiny{$-$0.070}}}$ \\
26.10--28.73 & 27.38 & $1.475 \pm 0.008 ^{\raisebox{-0.2ex}[0ex][-1ex]{\hspace{2.5mm}\tiny{$+$0.029}}}_{\raisebox{0.1ex}{\hspace{2.4mm}\tiny{$-$0.050}}}$ & $1.524 \pm 0.003 ^{\raisebox{-0.2ex}[0ex][-1ex]{\hspace{2.5mm}\tiny{$+$0.033}}}_{\raisebox{0.1ex}{\hspace{2.4mm}\tiny{$-$0.054}}}$ \\
28.73--31.62 & 30.13 & $1.137 \pm 0.007 ^{\raisebox{-0.2ex}[0ex][-1ex]{\hspace{2.5mm}\tiny{$+$0.022}}}_{\raisebox{0.1ex}{\hspace{2.4mm}\tiny{$-$0.038}}}$ & $1.177 \pm 0.003 ^{\raisebox{-0.2ex}[0ex][-1ex]{\hspace{2.5mm}\tiny{$+$0.026}}}_{\raisebox{0.1ex}{\hspace{2.4mm}\tiny{$-$0.042}}}$ \\
31.62--35.48 & 33.48 & $8.629 \pm 0.051 ^{\raisebox{-0.2ex}[0ex][-1ex]{\hspace{2.5mm}\tiny{$+$0.171}}}_{\raisebox{0.1ex}{\hspace{2.4mm}\tiny{$-$0.292}}} \times 10^{-1}$ & $8.816 \pm 0.019 ^{\raisebox{-0.2ex}[0ex][-1ex]{\hspace{2.5mm}\tiny{$+$0.195}}}_{\raisebox{0.1ex}{\hspace{2.4mm}\tiny{$-$0.318}}} \times 10^{-1}$ \\
35.48--39.81 & 37.57 & $6.254 \pm 0.041 ^{\raisebox{-0.2ex}[0ex][-1ex]{\hspace{2.5mm}\tiny{$+$0.128}}}_{\raisebox{0.1ex}{\hspace{2.4mm}\tiny{$-$0.214}}} \times 10^{-1}$ & $6.472 \pm 0.016 ^{\raisebox{-0.2ex}[0ex][-1ex]{\hspace{2.5mm}\tiny{$+$0.147}}}_{\raisebox{0.1ex}{\hspace{2.4mm}\tiny{$-$0.236}}} \times 10^{-1}$ \\
39.81--44.67 & 42.15 & $4.581 \pm 0.033 ^{\raisebox{-0.2ex}[0ex][-1ex]{\hspace{2.5mm}\tiny{$+$0.096}}}_{\raisebox{0.1ex}{\hspace{2.4mm}\tiny{$-$0.158}}} \times 10^{-1}$ & $4.707 \pm 0.013 ^{\raisebox{-0.2ex}[0ex][-1ex]{\hspace{2.5mm}\tiny{$+$0.108}}}_{\raisebox{0.1ex}{\hspace{2.4mm}\tiny{$-$0.172}}} \times 10^{-1}$ \\
44.67--50.12 & 47.30 & $3.350 \pm 0.027 ^{\raisebox{-0.2ex}[0ex][-1ex]{\hspace{2.5mm}\tiny{$+$0.073}}}_{\raisebox{0.1ex}{\hspace{2.4mm}\tiny{$-$0.118}}} \times 10^{-1}$ & $3.434 \pm 0.010 ^{\raisebox{-0.2ex}[0ex][-1ex]{\hspace{2.5mm}\tiny{$+$0.082}}}_{\raisebox{0.1ex}{\hspace{2.4mm}\tiny{$-$0.128}}} \times 10^{-1}$ \\
50.12--56.23 & 53.07 & $2.471 \pm 0.022 ^{\raisebox{-0.2ex}[0ex][-1ex]{\hspace{2.5mm}\tiny{$+$0.056}}}_{\raisebox{0.1ex}{\hspace{2.4mm}\tiny{$-$0.088}}} \times 10^{-1}$ & $2.491 \pm 0.008 ^{\raisebox{-0.2ex}[0ex][-1ex]{\hspace{2.5mm}\tiny{$+$0.061}}}_{\raisebox{0.1ex}{\hspace{2.4mm}\tiny{$-$0.093}}} \times 10^{-1}$ \\
56.23--63.10 & 59.54 & $1.791 \pm 0.017 ^{\raisebox{-0.2ex}[0ex][-1ex]{\hspace{2.5mm}\tiny{$+$0.043}}}_{\raisebox{0.1ex}{\hspace{2.4mm}\tiny{$-$0.065}}} \times 10^{-1}$ & $1.823 \pm 0.007 ^{\raisebox{-0.2ex}[0ex][-1ex]{\hspace{2.5mm}\tiny{$+$0.047}}}_{\raisebox{0.1ex}{\hspace{2.4mm}\tiny{$-$0.070}}} \times 10^{-1}$ \\
63.10--70.79 & 66.81 & $1.303 \pm 0.014 ^{\raisebox{-0.2ex}[0ex][-1ex]{\hspace{2.5mm}\tiny{$+$0.032}}}_{\raisebox{0.1ex}{\hspace{2.4mm}\tiny{$-$0.048}}} \times 10^{-1}$ & $1.327 \pm 0.005 ^{\raisebox{-0.2ex}[0ex][-1ex]{\hspace{2.5mm}\tiny{$+$0.035}}}_{\raisebox{0.1ex}{\hspace{2.4mm}\tiny{$-$0.052}}} \times 10^{-1}$ \\
70.79--79.43 & 74.96 & $9.673 \pm 0.114 ^{\raisebox{-0.2ex}[0ex][-1ex]{\hspace{2.5mm}\tiny{$+$0.260}}}_{\raisebox{0.1ex}{\hspace{2.4mm}\tiny{$-$0.371}}} \times 10^{-2}$ & $9.681 \pm 0.043 ^{\raisebox{-0.2ex}[0ex][-1ex]{\hspace{2.5mm}\tiny{$+$0.276}}}_{\raisebox{0.1ex}{\hspace{2.4mm}\tiny{$-$0.390}}} \times 10^{-2}$ \\
79.43--89.13 & 84.10 & $6.981 \pm 0.092 ^{\raisebox{-0.2ex}[0ex][-1ex]{\hspace{2.5mm}\tiny{$+$0.196}}}_{\raisebox{0.1ex}{\hspace{2.4mm}\tiny{$-$0.274}}} \times 10^{-2}$ & $7.055 \pm 0.035 ^{\raisebox{-0.2ex}[0ex][-1ex]{\hspace{2.5mm}\tiny{$+$0.209}}}_{\raisebox{0.1ex}{\hspace{2.4mm}\tiny{$-$0.289}}} \times 10^{-2}$ \\
89.13--100.0 & 94.36 & $4.959 \pm 0.073 ^{\raisebox{-0.2ex}[0ex][-1ex]{\hspace{2.5mm}\tiny{$+$0.153}}}_{\raisebox{0.1ex}{\hspace{2.4mm}\tiny{$-$0.205}}} \times 10^{-2}$ & $5.165 \pm 0.028 ^{\raisebox{-0.2ex}[0ex][-1ex]{\hspace{2.5mm}\tiny{$+$0.167}}}_{\raisebox{0.1ex}{\hspace{2.4mm}\tiny{$-$0.222}}} \times 10^{-2}$ \\
100.0--113.6 & 106.5 & $3.779 \pm 0.059 ^{\raisebox{-0.2ex}[0ex][-1ex]{\hspace{2.5mm}\tiny{$+$0.128}}}_{\raisebox{0.1ex}{\hspace{2.4mm}\tiny{$-$0.168}}} \times 10^{-2}$ & $3.736 \pm 0.023 ^{\raisebox{-0.2ex}[0ex][-1ex]{\hspace{2.5mm}\tiny{$+$0.133}}}_{\raisebox{0.1ex}{\hspace{2.4mm}\tiny{$-$0.174}}} \times 10^{-2}$ \\
113.6--129.2 & 121.1 & $2.639 \pm 0.046 ^{\raisebox{-0.2ex}[0ex][-1ex]{\hspace{2.5mm}\tiny{$+$0.101}}}_{\raisebox{0.1ex}{\hspace{2.4mm}\tiny{$-$0.125}}} \times 10^{-2}$ & $2.605 \pm 0.018 ^{\raisebox{-0.2ex}[0ex][-1ex]{\hspace{2.5mm}\tiny{$+$0.104}}}_{\raisebox{0.1ex}{\hspace{2.4mm}\tiny{$-$0.130}}} \times 10^{-2}$ \\
129.2--146.8 & 137.6 & $1.861 \pm 0.037 ^{\raisebox{-0.2ex}[0ex][-1ex]{\hspace{2.5mm}\tiny{$+$0.080}}}_{\raisebox{0.1ex}{\hspace{2.4mm}\tiny{$-$0.096}}} \times 10^{-2}$ & $1.881 \pm 0.014 ^{\raisebox{-0.2ex}[0ex][-1ex]{\hspace{2.5mm}\tiny{$+$0.084}}}_{\raisebox{0.1ex}{\hspace{2.4mm}\tiny{$-$0.101}}} \times 10^{-2}$ \\
146.8--166.8 & 156.4 & $1.322 \pm 0.029 ^{\raisebox{-0.2ex}[0ex][-1ex]{\hspace{2.5mm}\tiny{$+$0.061}}}_{\raisebox{0.1ex}{\hspace{2.4mm}\tiny{$-$0.072}}} \times 10^{-2}$ & $1.329 \pm 0.011 ^{\raisebox{-0.2ex}[0ex][-1ex]{\hspace{2.5mm}\tiny{$+$0.064}}}_{\raisebox{0.1ex}{\hspace{2.4mm}\tiny{$-$0.075}}} \times 10^{-2}$ \\
\enddata

%% file: table02.tex
\startdata
0.150--0.159 & 0.154 & $1.346 \pm 0.018 \pm 0.037 \times 10^{2}$ & $3.241 \pm 0.009 \pm 0.082 \times 10^{2}$ \\
0.159--0.168 & 0.163 & $1.406 \pm 0.018 \pm 0.038 \times 10^{2}$ & $3.312 \pm 0.009 \pm 0.084 \times 10^{2}$ \\
0.168--0.178 & 0.173 & $1.470 \pm 0.017 \pm 0.039 \times 10^{2}$ & $3.367 \pm 0.009 \pm 0.086 \times 10^{2}$ \\
0.178--0.188 & 0.183 & $1.540 \pm 0.017 \pm 0.041 \times 10^{2}$ & $3.431 \pm 0.009 \pm 0.087 \times 10^{2}$ \\
0.188--0.200 & 0.194 & $1.601 \pm 0.017 \pm 0.042 \times 10^{2}$ & $3.440 \pm 0.008 \pm 0.088 \times 10^{2}$ \\
0.200--0.211 & 0.205 & $1.583 \pm 0.016 \pm 0.041 \times 10^{2}$ & $3.429 \pm 0.008 \pm 0.088 \times 10^{2}$ \\
0.211--0.224 & 0.218 & $1.662 \pm 0.016 \pm 0.043 \times 10^{2}$ & $3.398 \pm 0.008 \pm 0.088 \times 10^{2}$ \\
0.224--0.237 & 0.231 & $1.654 \pm 0.015 \pm 0.042 \times 10^{2}$ & $3.372 \pm 0.007 \pm 0.087 \times 10^{2}$ \\
0.237--0.251 & 0.244 & $1.674 \pm 0.015 \pm 0.043 \times 10^{2}$ & $3.308 \pm 0.007 \pm 0.086 \times 10^{2}$ \\
0.251--0.266 & 0.259 & $1.706 \pm 0.014 \pm 0.043 \times 10^{2}$ & $3.245 \pm 0.007 \pm 0.085 \times 10^{2}$ \\
0.266--0.282 & 0.274 & $1.676 \pm 0.014 \pm 0.042 \times 10^{2}$ & $3.184 \pm 0.007 \pm 0.083 \times 10^{2}$ \\
0.282--0.298 & 0.290 & $1.675 \pm 0.013 \pm 0.042 \times 10^{2}$ & $3.080 \pm 0.006 \pm 0.081 \times 10^{2}$ \\
0.298--0.316 & 0.307 & $1.646 \pm 0.013 \pm 0.042 \times 10^{2}$ & $3.016 \pm 0.006 \pm 0.079 \times 10^{2}$ \\
0.316--0.335 & 0.326 & $1.630 \pm 0.012 \pm 0.041 \times 10^{2}$ & $2.911 \pm 0.006 \pm 0.076 \times 10^{2}$ \\
0.335--0.355 & 0.345 & $1.605 \pm 0.012 \pm 0.041 \times 10^{2}$ & $2.818 \pm 0.005 \pm 0.074 \times 10^{2}$ \\
0.355--0.376 & 0.365 & $1.604 \pm 0.011 \pm 0.041 \times 10^{2}$ & $2.731 \pm 0.005 \pm 0.072 \times 10^{2}$ \\
0.376--0.398 & 0.387 & $1.573 \pm 0.011 \pm 0.040 \times 10^{2}$ & $2.620 \pm 0.005 \pm 0.069 \times 10^{2}$ \\
0.398--0.422 & 0.410 & $1.531 \pm 0.010 \pm 0.039 \times 10^{2}$ & $2.530 \pm 0.005 \pm 0.067 \times 10^{2}$ \\
0.422--0.447 & 0.434 & $1.498 \pm 0.010 \pm 0.038 \times 10^{2}$ & $2.416 \pm 0.004 \pm 0.065 \times 10^{2}$ \\
0.447--0.473 & 0.460 & $1.467 \pm 0.009 \pm 0.038 \times 10^{2}$ & $2.319 \pm 0.004 \pm 0.062 \times 10^{2}$ \\
0.473--0.501 & 0.487 & $1.425 \pm 0.009 \pm 0.037 \times 10^{2}$ & $2.217 \pm 0.004 \pm 0.060 \times 10^{2}$ \\
0.501--0.531 & 0.516 & $1.382 \pm 0.009 \pm 0.036 \times 10^{2}$ & $2.110 \pm 0.004 \pm 0.058 \times 10^{2}$ \\
0.531--0.562 & 0.546 & $1.338 \pm 0.008 \pm 0.035 \times 10^{2}$ & $2.010 \pm 0.004 \pm 0.056 \times 10^{2}$ \\
0.562--0.596 & 0.579 & $1.272 \pm 0.008 \pm 0.034 \times 10^{2}$ & $1.908 \pm 0.003 \pm 0.053 \times 10^{2}$ \\
0.596--0.631 & 0.613 & $1.230 \pm 0.007 \pm 0.033 \times 10^{2}$ & $1.805 \pm 0.003 \pm 0.051 \times 10^{2}$ \\
0.631--0.668 & 0.649 & $1.195 \pm 0.007 \pm 0.032 \times 10^{2}$ & $1.701 \pm 0.003 \pm 0.049 \times 10^{2}$ \\
0.668--0.708 & 0.688 & $1.133 \pm 0.007 \pm 0.031 \times 10^{2}$ & $1.604 \pm 0.003 \pm 0.047 \times 10^{2}$ \\
0.708--0.750 & 0.729 & $1.079 \pm 0.006 \pm 0.030 \times 10^{2}$ & $1.511 \pm 0.003 \pm 0.044 \times 10^{2}$ \\
0.750--0.794 & 0.772 & $1.026 \pm 0.006 \pm 0.028 \times 10^{2}$ & $1.419 \pm 0.003 \pm 0.042 \times 10^{2}$ \\
0.794--0.841 & 0.818 & $9.866 \pm 0.055 \pm 0.273 \times 10^{1}$ & $1.326 \pm 0.002 \pm 0.039 \times 10^{2}$ \\
0.841--0.891 & 0.866 & $9.291 \pm 0.052 \pm 0.258 \times 10^{1}$ & $1.240 \pm 0.002 \pm 0.037 \times 10^{2}$ \\
0.891--0.944 & 0.917 & $8.909 \pm 0.050 \pm 0.248 \times 10^{1}$ & $1.153 \pm 0.002 \pm 0.034 \times 10^{2}$ \\
0.944--1.000 & 0.972 & $8.162 \pm 0.046 \pm 0.228 \times 10^{1}$ & $1.071 \pm 0.002 \pm 0.032 \times 10^{2}$ \\
1.000--1.080 & 1.039 & $7.661 \pm 0.037 \pm 0.214 \times 10^{1}$ & $9.850 \pm 0.016 \pm 0.292 \times 10^{1}$ \\
1.080--1.166 & 1.122 & $6.971 \pm 0.034 \pm 0.195 \times 10^{1}$ & $8.855 \pm 0.014 \pm 0.262 \times 10^{1}$ \\
1.166--1.259 & 1.212 & $6.444 \pm 0.032 \pm 0.180 \times 10^{1}$ & $7.929 \pm 0.013 \pm 0.234 \times 10^{1}$ \\
1.259--1.359 & 1.308 & $5.773 \pm 0.029 \pm 0.161 \times 10^{1}$ & $7.099 \pm 0.012 \pm 0.210 \times 10^{1}$ \\
1.359--1.468 & 1.413 & $5.209 \pm 0.027 \pm 0.145 \times 10^{1}$ & $6.342 \pm 0.011 \pm 0.187 \times 10^{1}$ \\
1.468--1.585 & 1.525 & $4.724 \pm 0.025 \pm 0.132 \times 10^{1}$ & $5.613 \pm 0.010 \pm 0.166 \times 10^{1}$ \\
1.585--1.711 & 1.647 & $4.188 \pm 0.022 \pm 0.116 \times 10^{1}$ & $4.962 \pm 0.009 \pm 0.146 \times 10^{1}$ \\
1.711--1.848 & 1.778 & $3.749 \pm 0.020 \pm 0.104 \times 10^{1}$ & $4.360 \pm 0.008 \pm 0.128 \times 10^{1}$ \\
1.848--1.995 & 1.920 & $3.331 \pm 0.018 \pm 0.092 \times 10^{1}$ & $3.840 \pm 0.007 \pm 0.113 \times 10^{1}$ \\
1.995--2.154 & 2.073 & $2.944 \pm 0.017 \pm 0.081 \times 10^{1}$ & $3.376 \pm 0.007 \pm 0.099 \times 10^{1}$ \\
2.154--2.326 & 2.239 & $2.616 \pm 0.015 \pm 0.071 \times 10^{1}$ & $2.953 \pm 0.006 \pm 0.086 \times 10^{1}$ \\
2.326--2.512 & 2.417 & $2.274 \pm 0.014 \pm 0.062 \times 10^{1}$ & $2.572 \pm 0.005 \pm 0.075 \times 10^{1}$ \\
2.512--2.712 & 2.610 & $2.004 \pm 0.012 \pm 0.054 \times 10^{1}$ & $2.227 \pm 0.005 \pm 0.065 \times 10^{1}$ \\
2.712--2.929 & 2.818 & $1.759 \pm 0.011 \pm 0.047 \times 10^{1}$ & $1.936 \pm 0.004 \pm 0.056 \times 10^{1}$ \\
2.929--3.162 & 3.043 & $1.522 \pm 0.010 \pm 0.041 \times 10^{1}$ & $1.671 \pm 0.004 \pm 0.048 \times 10^{1}$ \\
3.162--3.481 & 3.318 & $1.297 \pm 0.008 \pm 0.035 \times 10^{1}$ & $1.414 \pm 0.003 \pm 0.041 \times 10^{1}$ \\
3.481--3.831 & 3.652 & $1.077 \pm 0.007 \pm 0.029 \times 10^{1}$ & $1.167 \pm 0.003 \pm 0.034 \times 10^{1}$ \\
3.831--4.217 & 4.019 & $8.975 \pm 0.060 \pm 0.236$ & $9.528 \pm 0.023 \pm 0.272$ \\
4.217--4.642 & 4.424 & $7.281 \pm 0.052 \pm 0.191$ & $7.807 \pm 0.020 \pm 0.223$ \\
4.642--5.109 & 4.869 & $6.050 \pm 0.045 \pm 0.159$ & $6.403 \pm 0.018 \pm 0.182$ \\
5.109--5.623 & 5.360 & $4.949 \pm 0.039 \pm 0.130$ & $5.190 \pm 0.015 \pm 0.148$ \\
5.623--6.190 & 5.899 & $3.934 \pm 0.033 \pm 0.103$ & $4.157 \pm 0.013 \pm 0.118$ \\
6.190--6.813 & 6.493 & $3.228 \pm 0.029 \pm 0.085$ & $3.355 \pm 0.011 \pm 0.096$ \\
6.813--7.499 & 7.147 & $2.574 \pm 0.025 \pm 0.068$ & $2.700 \pm 0.009 \pm 0.077$ \\
7.499--8.254 & 7.866 & $2.044 \pm 0.021 \pm 0.054$ & $2.159 \pm 0.008 \pm 0.061$ \\
8.254--9.085 & 8.658 & $1.675 \pm 0.018 \pm 0.045$ & $1.727 \pm 0.007 \pm 0.049$ \\
9.085--10.00 & 9.530 & $1.288 \pm 0.015 \pm 0.035$ & $1.355 \pm 0.006 \pm 0.039$ \\
10.00--11.22 & 10.59 & $1.020 \pm 0.012 \pm 0.028$ & $1.061 \pm 0.004 \pm 0.030$ \\
11.22--12.59 & 11.88 & $7.776 \pm 0.096 \pm 0.216 \times 10^{-1}$ & $7.912 \pm 0.036 \pm 0.228 \times 10^{-1}$ \\
12.59--14.13 & 13.33 & $5.714 \pm 0.078 \pm 0.162 \times 10^{-1}$ & $5.970 \pm 0.030 \pm 0.172 \times 10^{-1}$ \\
14.13--15.85 & 14.96 & $4.305 \pm 0.064 \pm 0.125 \times 10^{-1}$ & $4.478 \pm 0.025 \pm 0.130 \times 10^{-1}$ \\
15.85--17.78 & 16.78 & $3.230 \pm 0.052 \pm 0.096 \times 10^{-1}$ & $3.330 \pm 0.020 \pm 0.097 \times 10^{-1}$ \\
17.78--19.95 & 18.83 & $2.415 \pm 0.043 \pm 0.075 \times 10^{-1}$ & $2.498 \pm 0.016 \pm 0.074 \times 10^{-1}$ \\
19.95--22.39 & 21.13 & $1.697 \pm 0.034 \pm 0.054 \times 10^{-1}$ & $1.826 \pm 0.013 \pm 0.055 \times 10^{-1}$ \\
22.39--25.12 & 23.71 & $1.322 \pm 0.028 \pm 0.044 \times 10^{-1}$ & $1.377 \pm 0.011 \pm 0.042 \times 10^{-1}$ \\
25.12--28.18 & 26.60 & $9.943 \pm 0.234 \pm 0.343 \times 10^{-2}$ & $1.004 \pm 0.009 \pm 0.031 \times 10^{-1}$ \\
28.18--31.62 & 29.84 & $7.177 \pm 0.188 \pm 0.258 \times 10^{-2}$ & $7.565 \pm 0.072 \pm 0.240 \times 10^{-2}$ \\
31.62--38.31 & 34.77 & $4.778 \pm 0.110 \pm 0.181 \times 10^{-2}$ & $4.901 \pm 0.042 \pm 0.161 \times 10^{-2}$ \\
38.31--46.42 & 42.12 & $2.837 \pm 0.077 \pm 0.113 \times 10^{-2}$ & $2.950 \pm 0.029 \pm 0.101 \times 10^{-2}$ \\
46.42--56.23 & 51.03 & $1.750 \pm 0.055 \pm 0.074 \times 10^{-2}$ & $1.774 \pm 0.021 \pm 0.066 \times 10^{-2}$ \\
56.23--68.13 & 61.83 & $1.121 \pm 0.040 \pm 0.049 \times 10^{-2}$ & $1.109 \pm 0.015 \pm 0.044 \times 10^{-2}$ \\
68.13--82.54 & 74.91 & $6.627 \pm 0.283 \pm 0.320 \times 10^{-3}$ & $6.546 \pm 0.104 \pm 0.293 \times 10^{-3}$ \\
\enddata

%% file: table03.tex
\startdata
1.098--1.132 & 1.115 & $1.638 \pm 0.022 \pm 0.056 \times 10^{1}$ & $1.452 \pm 0.004 \pm 0.051 \times 10^{1}$ \\
1.132--1.168 & 1.150 & $1.551 \pm 0.020 \pm 0.051 \times 10^{1}$ & $1.375 \pm 0.004 \pm 0.048 \times 10^{1}$ \\
1.168--1.205 & 1.187 & $1.461 \pm 0.018 \pm 0.048 \times 10^{1}$ & $1.306 \pm 0.003 \pm 0.046 \times 10^{1}$ \\
1.205--1.244 & 1.224 & $1.374 \pm 0.015 \pm 0.044 \times 10^{1}$ & $1.237 \pm 0.003 \pm 0.042 \times 10^{1}$ \\
1.244--1.283 & 1.263 & $1.294 \pm 0.014 \pm 0.040 \times 10^{1}$ & $1.189 \pm 0.003 \pm 0.041 \times 10^{1}$ \\
1.283--1.325 & 1.304 & $1.281 \pm 0.013 \pm 0.039 \times 10^{1}$ & $1.145 \pm 0.003 \pm 0.039 \times 10^{1}$ \\
1.325--1.367 & 1.346 & $1.191 \pm 0.012 \pm 0.036 \times 10^{1}$ & $1.107 \pm 0.003 \pm 0.037 \times 10^{1}$ \\
1.367--1.412 & 1.389 & $1.168 \pm 0.011 \pm 0.034 \times 10^{1}$ & $1.068 \pm 0.002 \pm 0.036 \times 10^{1}$ \\
1.412--1.458 & 1.435 & $1.123 \pm 0.010 \pm 0.033 \times 10^{1}$ & $1.039 \pm 0.002 \pm 0.035 \times 10^{1}$ \\
1.458--1.506 & 1.482 & $1.075 \pm 0.009 \pm 0.031 \times 10^{1}$ & $1.012 \pm 0.002 \pm 0.033 \times 10^{1}$ \\
1.506--1.555 & 1.530 & $1.060 \pm 0.009 \pm 0.030 \times 10^{1}$ & $9.805 \pm 0.021 \pm 0.322$ \\
1.555--1.607 & 1.581 & $1.024 \pm 0.008 \pm 0.029 \times 10^{1}$ & $9.634 \pm 0.020 \pm 0.312$ \\
1.607--1.661 & 1.634 & $1.008 \pm 0.008 \pm 0.028 \times 10^{1}$ & $9.326 \pm 0.019 \pm 0.298$ \\
1.661--1.717 & 1.688 & $9.780 \pm 0.075 \pm 0.269$ & $9.133 \pm 0.019 \pm 0.292$ \\
1.717--1.774 & 1.745 & $9.557 \pm 0.072 \pm 0.259$ & $8.915 \pm 0.018 \pm 0.282$ \\
1.774--1.835 & 1.805 & $9.150 \pm 0.066 \pm 0.246$ & $8.670 \pm 0.017 \pm 0.272$ \\
1.835--1.898 & 1.866 & $8.907 \pm 0.063 \pm 0.240$ & $8.510 \pm 0.017 \pm 0.267$ \\
1.898--1.964 & 1.930 & $8.797 \pm 0.061 \pm 0.237$ & $8.295 \pm 0.016 \pm 0.260$ \\
1.964--2.032 & 1.997 & $8.544 \pm 0.058 \pm 0.230$ & $8.147 \pm 0.016 \pm 0.256$ \\
2.032--2.103 & 2.067 & $8.319 \pm 0.056 \pm 0.225$ & $7.961 \pm 0.015 \pm 0.250$ \\
2.103--2.178 & 2.140 & $8.167 \pm 0.053 \pm 0.222$ & $7.793 \pm 0.015 \pm 0.247$ \\
2.178--2.255 & 2.216 & $7.945 \pm 0.051 \pm 0.217$ & $7.657 \pm 0.014 \pm 0.245$ \\
2.255--2.336 & 2.295 & $7.795 \pm 0.049 \pm 0.217$ & $7.505 \pm 0.014 \pm 0.244$ \\
2.336--2.421 & 2.378 & $7.760 \pm 0.049 \pm 0.215$ & $7.376 \pm 0.014 \pm 0.238$ \\
2.421--2.510 & 2.465 & $7.580 \pm 0.047 \pm 0.213$ & $7.267 \pm 0.013 \pm 0.239$ \\
2.510--2.601 & 2.555 & $7.338 \pm 0.045 \pm 0.209$ & $7.171 \pm 0.013 \pm 0.239$ \\
2.601--2.699 & 2.650 & $7.277 \pm 0.044 \pm 0.210$ & $7.050 \pm 0.013 \pm 0.238$ \\
2.699--2.800 & 2.749 & $7.190 \pm 0.043 \pm 0.207$ & $6.933 \pm 0.013 \pm 0.234$ \\
2.800--2.907 & 2.853 & $7.059 \pm 0.043 \pm 0.206$ & $6.836 \pm 0.013 \pm 0.233$ \\
2.907--3.018 & 2.961 & $6.882 \pm 0.041 \pm 0.202$ & $6.751 \pm 0.013 \pm 0.232$ \\
3.018--3.134 & 3.075 & $6.854 \pm 0.041 \pm 0.202$ & $6.664 \pm 0.013 \pm 0.230$ \\
3.134--3.256 & 3.194 & $6.692 \pm 0.040 \pm 0.197$ & $6.602 \pm 0.013 \pm 0.228$ \\
3.256--3.385 & 3.319 & $6.788 \pm 0.041 \pm 0.201$ & $6.547 \pm 0.013 \pm 0.226$ \\
3.385--3.566 & 3.474 & $6.668 \pm 0.035 \pm 0.196$ & $6.473 \pm 0.011 \pm 0.223$ \\
3.566--3.759 & 3.661 & $6.612 \pm 0.035 ^{\raisebox{-0.2ex}[0ex][-1ex]{\hspace{2.5mm}\tiny{$+$0.194}}}_{\raisebox{0.1ex}{\hspace{2.4mm}\tiny{$-$0.195}}}$ & $6.426 \pm 0.011 ^{\raisebox{-0.2ex}[0ex][-1ex]{\hspace{2.5mm}\tiny{$+$0.220}}}_{\raisebox{0.1ex}{\hspace{2.4mm}\tiny{$-$0.221}}}$ \\
3.759--3.966 & 3.861 & $6.499 \pm 0.034 ^{\raisebox{-0.2ex}[0ex][-1ex]{\hspace{2.5mm}\tiny{$+$0.190}}}_{\raisebox{0.1ex}{\hspace{2.4mm}\tiny{$-$0.192}}}$ & $6.397 \pm 0.011 ^{\raisebox{-0.2ex}[0ex][-1ex]{\hspace{2.5mm}\tiny{$+$0.219}}}_{\raisebox{0.1ex}{\hspace{2.4mm}\tiny{$-$0.221}}}$ \\
3.966--4.186 & 4.075 & $6.492 \pm 0.035 ^{\raisebox{-0.2ex}[0ex][-1ex]{\hspace{2.5mm}\tiny{$+$0.189}}}_{\raisebox{0.1ex}{\hspace{2.4mm}\tiny{$-$0.194}}}$ & $6.367 \pm 0.011 ^{\raisebox{-0.2ex}[0ex][-1ex]{\hspace{2.5mm}\tiny{$+$0.217}}}_{\raisebox{0.1ex}{\hspace{2.4mm}\tiny{$-$0.224}}}$ \\
4.186--4.423 & 4.303 & $6.418 \pm 0.035 ^{\raisebox{-0.2ex}[0ex][-1ex]{\hspace{2.5mm}\tiny{$+$0.186}}}_{\raisebox{0.1ex}{\hspace{2.4mm}\tiny{$-$0.200}}}$ & $6.308 \pm 0.011 ^{\raisebox{-0.2ex}[0ex][-1ex]{\hspace{2.5mm}\tiny{$+$0.214}}}_{\raisebox{0.1ex}{\hspace{2.4mm}\tiny{$-$0.233}}}$ \\
4.423--4.676 & 4.547 & $6.316 \pm 0.035 ^{\raisebox{-0.2ex}[0ex][-1ex]{\hspace{2.5mm}\tiny{$+$0.182}}}_{\raisebox{0.1ex}{\hspace{2.4mm}\tiny{$-$0.208}}}$ & $6.289 \pm 0.012 ^{\raisebox{-0.2ex}[0ex][-1ex]{\hspace{2.5mm}\tiny{$+$0.212}}}_{\raisebox{0.1ex}{\hspace{2.4mm}\tiny{$-$0.247}}}$ \\
4.676--4.947 & 4.810 & $6.315 \pm 0.035 ^{\raisebox{-0.2ex}[0ex][-1ex]{\hspace{2.5mm}\tiny{$+$0.181}}}_{\raisebox{0.1ex}{\hspace{2.4mm}\tiny{$-$0.207}}}$ & $6.242 \pm 0.012 ^{\raisebox{-0.2ex}[0ex][-1ex]{\hspace{2.5mm}\tiny{$+$0.210}}}_{\raisebox{0.1ex}{\hspace{2.4mm}\tiny{$-$0.244}}}$ \\
4.947--5.238 & 5.090 & $6.247 \pm 0.036 ^{\raisebox{-0.2ex}[0ex][-1ex]{\hspace{2.5mm}\tiny{$+$0.178}}}_{\raisebox{0.1ex}{\hspace{2.4mm}\tiny{$-$0.207}}}$ & $6.227 \pm 0.012 ^{\raisebox{-0.2ex}[0ex][-1ex]{\hspace{2.5mm}\tiny{$+$0.208}}}_{\raisebox{0.1ex}{\hspace{2.4mm}\tiny{$-$0.247}}}$ \\
5.238--5.550 & 5.391 & $6.173 \pm 0.036 ^{\raisebox{-0.2ex}[0ex][-1ex]{\hspace{2.5mm}\tiny{$+$0.175}}}_{\raisebox{0.1ex}{\hspace{2.4mm}\tiny{$-$0.210}}}$ & $6.169 \pm 0.013 ^{\raisebox{-0.2ex}[0ex][-1ex]{\hspace{2.5mm}\tiny{$+$0.205}}}_{\raisebox{0.1ex}{\hspace{2.4mm}\tiny{$-$0.249}}}$ \\
5.550--5.885 & 5.715 & $6.152 \pm 0.037 ^{\raisebox{-0.2ex}[0ex][-1ex]{\hspace{2.5mm}\tiny{$+$0.173}}}_{\raisebox{0.1ex}{\hspace{2.4mm}\tiny{$-$0.215}}}$ & $6.120 \pm 0.013 ^{\raisebox{-0.2ex}[0ex][-1ex]{\hspace{2.5mm}\tiny{$+$0.203}}}_{\raisebox{0.1ex}{\hspace{2.4mm}\tiny{$-$0.254}}}$ \\
5.885--6.244 & 6.061 & $6.047 \pm 0.037 ^{\raisebox{-0.2ex}[0ex][-1ex]{\hspace{2.5mm}\tiny{$+$0.169}}}_{\raisebox{0.1ex}{\hspace{2.4mm}\tiny{$-$0.220}}}$ & $6.065 \pm 0.013 ^{\raisebox{-0.2ex}[0ex][-1ex]{\hspace{2.5mm}\tiny{$+$0.200}}}_{\raisebox{0.1ex}{\hspace{2.4mm}\tiny{$-$0.259}}}$ \\
6.244--6.630 & 6.434 & $6.060 \pm 0.039 ^{\raisebox{-0.2ex}[0ex][-1ex]{\hspace{2.5mm}\tiny{$+$0.168}}}_{\raisebox{0.1ex}{\hspace{2.4mm}\tiny{$-$0.226}}}$ & $6.033 \pm 0.013 ^{\raisebox{-0.2ex}[0ex][-1ex]{\hspace{2.5mm}\tiny{$+$0.198}}}_{\raisebox{0.1ex}{\hspace{2.4mm}\tiny{$-$0.260}}}$ \\
6.630--7.046 & 6.834 & $5.968 \pm 0.039 ^{\raisebox{-0.2ex}[0ex][-1ex]{\hspace{2.5mm}\tiny{$+$0.165}}}_{\raisebox{0.1ex}{\hspace{2.4mm}\tiny{$-$0.228}}}$ & $6.003 \pm 0.014 ^{\raisebox{-0.2ex}[0ex][-1ex]{\hspace{2.5mm}\tiny{$+$0.196}}}_{\raisebox{0.1ex}{\hspace{2.4mm}\tiny{$-$0.262}}}$ \\
7.046--7.493 & 7.265 & $5.914 \pm 0.040 ^{\raisebox{-0.2ex}[0ex][-1ex]{\hspace{2.5mm}\tiny{$+$0.162}}}_{\raisebox{0.1ex}{\hspace{2.4mm}\tiny{$-$0.226}}}$ & $5.943 \pm 0.014 ^{\raisebox{-0.2ex}[0ex][-1ex]{\hspace{2.5mm}\tiny{$+$0.194}}}_{\raisebox{0.1ex}{\hspace{2.4mm}\tiny{$-$0.260}}}$ \\
7.493--7.973 & 7.729 & $5.866 \pm 0.041 ^{\raisebox{-0.2ex}[0ex][-1ex]{\hspace{2.5mm}\tiny{$+$0.160}}}_{\raisebox{0.1ex}{\hspace{2.4mm}\tiny{$-$0.225}}}$ & $5.897 \pm 0.015 ^{\raisebox{-0.2ex}[0ex][-1ex]{\hspace{2.5mm}\tiny{$+$0.191}}}_{\raisebox{0.1ex}{\hspace{2.4mm}\tiny{$-$0.258}}}$ \\
7.973--8.626 & 8.292 & $5.802 \pm 0.038 ^{\raisebox{-0.2ex}[0ex][-1ex]{\hspace{2.5mm}\tiny{$+$0.157}}}_{\raisebox{0.1ex}{\hspace{2.4mm}\tiny{$-$0.226}}}$ & $5.832 \pm 0.014 ^{\raisebox{-0.2ex}[0ex][-1ex]{\hspace{2.5mm}\tiny{$+$0.188}}}_{\raisebox{0.1ex}{\hspace{2.4mm}\tiny{$-$0.256}}}$ \\
8.626--9.342 & 8.975 & $5.776 \pm 0.040 ^{\raisebox{-0.2ex}[0ex][-1ex]{\hspace{2.5mm}\tiny{$+$0.155}}}_{\raisebox{0.1ex}{\hspace{2.4mm}\tiny{$-$0.226}}}$ & $5.778 \pm 0.014 ^{\raisebox{-0.2ex}[0ex][-1ex]{\hspace{2.5mm}\tiny{$+$0.185}}}_{\raisebox{0.1ex}{\hspace{2.4mm}\tiny{$-$0.254}}}$ \\
9.342--10.13 & 9.725 & $5.663 \pm 0.041 ^{\raisebox{-0.2ex}[0ex][-1ex]{\hspace{2.5mm}\tiny{$+$0.151}}}_{\raisebox{0.1ex}{\hspace{2.4mm}\tiny{$-$0.221}}}$ & $5.744 \pm 0.015 ^{\raisebox{-0.2ex}[0ex][-1ex]{\hspace{2.5mm}\tiny{$+$0.183}}}_{\raisebox{0.1ex}{\hspace{2.4mm}\tiny{$-$0.251}}}$ \\
10.13--10.99 & 10.55 & $5.660 \pm 0.043 ^{\raisebox{-0.2ex}[0ex][-1ex]{\hspace{2.5mm}\tiny{$+$0.150}}}_{\raisebox{0.1ex}{\hspace{2.4mm}\tiny{$-$0.220}}}$ & $5.688 \pm 0.016 ^{\raisebox{-0.2ex}[0ex][-1ex]{\hspace{2.5mm}\tiny{$+$0.180}}}_{\raisebox{0.1ex}{\hspace{2.4mm}\tiny{$-$0.248}}}$ \\
10.99--11.94 & 11.45 & $5.540 \pm 0.044 ^{\raisebox{-0.2ex}[0ex][-1ex]{\hspace{2.5mm}\tiny{$+$0.146}}}_{\raisebox{0.1ex}{\hspace{2.4mm}\tiny{$-$0.215}}}$ & $5.584 \pm 0.016 ^{\raisebox{-0.2ex}[0ex][-1ex]{\hspace{2.5mm}\tiny{$+$0.177}}}_{\raisebox{0.1ex}{\hspace{2.4mm}\tiny{$-$0.243}}}$ \\
11.94--12.97 & 12.44 & $5.454 \pm 0.046 ^{\raisebox{-0.2ex}[0ex][-1ex]{\hspace{2.5mm}\tiny{$+$0.144}}}_{\raisebox{0.1ex}{\hspace{2.4mm}\tiny{$-$0.212}}}$ & $5.515 \pm 0.017 ^{\raisebox{-0.2ex}[0ex][-1ex]{\hspace{2.5mm}\tiny{$+$0.174}}}_{\raisebox{0.1ex}{\hspace{2.4mm}\tiny{$-$0.240}}}$ \\
12.97--14.12 & 13.54 & $5.505 \pm 0.050 ^{\raisebox{-0.2ex}[0ex][-1ex]{\hspace{2.5mm}\tiny{$+$0.146}}}_{\raisebox{0.1ex}{\hspace{2.4mm}\tiny{$-$0.214}}}$ & $5.525 \pm 0.018 ^{\raisebox{-0.2ex}[0ex][-1ex]{\hspace{2.5mm}\tiny{$+$0.175}}}_{\raisebox{0.1ex}{\hspace{2.4mm}\tiny{$-$0.241}}}$ \\
14.12--15.38 & 14.73 & $5.389 \pm 0.051 ^{\raisebox{-0.2ex}[0ex][-1ex]{\hspace{2.5mm}\tiny{$+$0.143}}}_{\raisebox{0.1ex}{\hspace{2.4mm}\tiny{$-$0.210}}}$ & $5.441 \pm 0.019 ^{\raisebox{-0.2ex}[0ex][-1ex]{\hspace{2.5mm}\tiny{$+$0.172}}}_{\raisebox{0.1ex}{\hspace{2.4mm}\tiny{$-$0.237}}}$ \\
15.38--16.76 & 16.05 & $5.342 \pm 0.054 ^{\raisebox{-0.2ex}[0ex][-1ex]{\hspace{2.5mm}\tiny{$+$0.142}}}_{\raisebox{0.1ex}{\hspace{2.4mm}\tiny{$-$0.209}}}$ & $5.368 \pm 0.020 ^{\raisebox{-0.2ex}[0ex][-1ex]{\hspace{2.5mm}\tiny{$+$0.170}}}_{\raisebox{0.1ex}{\hspace{2.4mm}\tiny{$-$0.234}}}$ \\
16.76--18.27 & 17.49 & $5.329 \pm 0.058 ^{\raisebox{-0.2ex}[0ex][-1ex]{\hspace{2.5mm}\tiny{$+$0.143}}}_{\raisebox{0.1ex}{\hspace{2.4mm}\tiny{$-$0.209}}}$ & $5.305 \pm 0.021 ^{\raisebox{-0.2ex}[0ex][-1ex]{\hspace{2.5mm}\tiny{$+$0.169}}}_{\raisebox{0.1ex}{\hspace{2.4mm}\tiny{$-$0.232}}}$ \\
18.27--19.95 & 19.09 & $5.208 \pm 0.060 ^{\raisebox{-0.2ex}[0ex][-1ex]{\hspace{2.5mm}\tiny{$+$0.141}}}_{\raisebox{0.1ex}{\hspace{2.4mm}\tiny{$-$0.205}}}$ & $5.225 \pm 0.022 ^{\raisebox{-0.2ex}[0ex][-1ex]{\hspace{2.5mm}\tiny{$+$0.167}}}_{\raisebox{0.1ex}{\hspace{2.4mm}\tiny{$-$0.229}}}$ \\
19.95--21.79 & 20.84 & $5.336 \pm 0.067 ^{\raisebox{-0.2ex}[0ex][-1ex]{\hspace{2.5mm}\tiny{$+$0.146}}}_{\raisebox{0.1ex}{\hspace{2.4mm}\tiny{$-$0.211}}}$ & $5.234 \pm 0.024 ^{\raisebox{-0.2ex}[0ex][-1ex]{\hspace{2.5mm}\tiny{$+$0.168}}}_{\raisebox{0.1ex}{\hspace{2.4mm}\tiny{$-$0.230}}}$ \\
21.79--24.23 & 22.97 & $5.125 \pm 0.063 ^{\raisebox{-0.2ex}[0ex][-1ex]{\hspace{2.5mm}\tiny{$+$0.143}}}_{\raisebox{0.1ex}{\hspace{2.4mm}\tiny{$-$0.205}}}$ & $5.121 \pm 0.023 ^{\raisebox{-0.2ex}[0ex][-1ex]{\hspace{2.5mm}\tiny{$+$0.165}}}_{\raisebox{0.1ex}{\hspace{2.4mm}\tiny{$-$0.226}}}$ \\
24.23--26.98 & 25.56 & $4.987 \pm 0.066 ^{\raisebox{-0.2ex}[0ex][-1ex]{\hspace{2.5mm}\tiny{$+$0.143}}}_{\raisebox{0.1ex}{\hspace{2.4mm}\tiny{$-$0.201}}}$ & $5.122 \pm 0.025 ^{\raisebox{-0.2ex}[0ex][-1ex]{\hspace{2.5mm}\tiny{$+$0.166}}}_{\raisebox{0.1ex}{\hspace{2.4mm}\tiny{$-$0.227}}}$ \\
26.98--30.05 & 28.46 & $5.094 \pm 0.074 ^{\raisebox{-0.2ex}[0ex][-1ex]{\hspace{2.5mm}\tiny{$+$0.150}}}_{\raisebox{0.1ex}{\hspace{2.4mm}\tiny{$-$0.209}}}$ & $5.037 \pm 0.027 ^{\raisebox{-0.2ex}[0ex][-1ex]{\hspace{2.5mm}\tiny{$+$0.165}}}_{\raisebox{0.1ex}{\hspace{2.4mm}\tiny{$-$0.224}}}$ \\
30.05--33.51 & 31.72 & $4.985 \pm 0.079 ^{\raisebox{-0.2ex}[0ex][-1ex]{\hspace{2.5mm}\tiny{$+$0.152}}}_{\raisebox{0.1ex}{\hspace{2.4mm}\tiny{$-$0.208}}}$ & $4.949 \pm 0.029 ^{\raisebox{-0.2ex}[0ex][-1ex]{\hspace{2.5mm}\tiny{$+$0.165}}}_{\raisebox{0.1ex}{\hspace{2.4mm}\tiny{$-$0.222}}}$ \\
33.51--37.39 & 35.38 & $4.962 \pm 0.086 ^{\raisebox{-0.2ex}[0ex][-1ex]{\hspace{2.5mm}\tiny{$+$0.157}}}_{\raisebox{0.1ex}{\hspace{2.4mm}\tiny{$-$0.212}}}$ & $4.909 \pm 0.032 ^{\raisebox{-0.2ex}[0ex][-1ex]{\hspace{2.5mm}\tiny{$+$0.166}}}_{\raisebox{0.1ex}{\hspace{2.4mm}\tiny{$-$0.222}}}$ \\
37.39--41.73 & 39.48 & $4.802 \pm 0.091 ^{\raisebox{-0.2ex}[0ex][-1ex]{\hspace{2.5mm}\tiny{$+$0.159}}}_{\raisebox{0.1ex}{\hspace{2.4mm}\tiny{$-$0.210}}}$ & $4.829 \pm 0.034 ^{\raisebox{-0.2ex}[0ex][-1ex]{\hspace{2.5mm}\tiny{$+$0.166}}}_{\raisebox{0.1ex}{\hspace{2.4mm}\tiny{$-$0.221}}}$ \\
41.73--46.60 & 44.08 & $5.074 \pm 0.108 ^{\raisebox{-0.2ex}[0ex][-1ex]{\hspace{2.5mm}\tiny{$+$0.177}}}_{\raisebox{0.1ex}{\hspace{2.4mm}\tiny{$-$0.228}}}$ & $4.830 \pm 0.038 ^{\raisebox{-0.2ex}[0ex][-1ex]{\hspace{2.5mm}\tiny{$+$0.170}}}_{\raisebox{0.1ex}{\hspace{2.4mm}\tiny{$-$0.224}}}$ \\
46.60--52.07 & 49.24 & $4.808 \pm 0.110 ^{\raisebox{-0.2ex}[0ex][-1ex]{\hspace{2.5mm}\tiny{$+$0.176}}}_{\raisebox{0.1ex}{\hspace{2.4mm}\tiny{$-$0.223}}}$ & $4.721 \pm 0.040 ^{\raisebox{-0.2ex}[0ex][-1ex]{\hspace{2.5mm}\tiny{$+$0.171}}}_{\raisebox{0.1ex}{\hspace{2.4mm}\tiny{$-$0.222}}}$ \\
52.07--58.20 & 55.03 & $4.721 \pm 0.119 ^{\raisebox{-0.2ex}[0ex][-1ex]{\hspace{2.5mm}\tiny{$+$0.181}}}_{\raisebox{0.1ex}{\hspace{2.4mm}\tiny{$-$0.225}}}$ & $4.707 \pm 0.044 ^{\raisebox{-0.2ex}[0ex][-1ex]{\hspace{2.5mm}\tiny{$+$0.174}}}_{\raisebox{0.1ex}{\hspace{2.4mm}\tiny{$-$0.224}}}$ \\
58.20--65.08 & 61.52 & $4.729 \pm 0.133 ^{\raisebox{-0.2ex}[0ex][-1ex]{\hspace{2.5mm}\tiny{$+$0.193}}}_{\raisebox{0.1ex}{\hspace{2.4mm}\tiny{$-$0.235}}}$ & $4.595 \pm 0.047 ^{\raisebox{-0.2ex}[0ex][-1ex]{\hspace{2.5mm}\tiny{$+$0.179}}}_{\raisebox{0.1ex}{\hspace{2.4mm}\tiny{$-$0.226}}}$ \\
65.08--78.46 & 71.37 & $4.772 \pm 0.118 ^{\raisebox{-0.2ex}[0ex][-1ex]{\hspace{2.5mm}\tiny{$+$0.209}}}_{\raisebox{0.1ex}{\hspace{2.4mm}\tiny{$-$0.249}}}$ & $4.680 \pm 0.043 ^{\raisebox{-0.2ex}[0ex][-1ex]{\hspace{2.5mm}\tiny{$+$0.193}}}_{\raisebox{0.1ex}{\hspace{2.4mm}\tiny{$-$0.239}}}$ \\
78.46--94.67 & 86.08 & $4.713 \pm 0.137 ^{\raisebox{-0.2ex}[0ex][-1ex]{\hspace{2.5mm}\tiny{$+$0.222}}}_{\raisebox{0.1ex}{\hspace{2.4mm}\tiny{$-$0.260}}}$ & $4.641 \pm 0.049 ^{\raisebox{-0.2ex}[0ex][-1ex]{\hspace{2.5mm}\tiny{$+$0.206}}}_{\raisebox{0.1ex}{\hspace{2.4mm}\tiny{$-$0.248}}}$ \\
94.67--114.3 & 103.9 & $4.646 \pm 0.157 ^{\raisebox{-0.2ex}[0ex][-1ex]{\hspace{2.5mm}\tiny{$+$0.245}}}_{\raisebox{0.1ex}{\hspace{2.4mm}\tiny{$-$0.278}}}$ & $4.599 \pm 0.058 ^{\raisebox{-0.2ex}[0ex][-1ex]{\hspace{2.5mm}\tiny{$+$0.231}}}_{\raisebox{0.1ex}{\hspace{2.4mm}\tiny{$-$0.269}}}$ \\
114.3--138.1 & 125.5 & $4.361 \pm 0.169 ^{\raisebox{-0.2ex}[0ex][-1ex]{\hspace{2.5mm}\tiny{$+$0.249}}}_{\raisebox{0.1ex}{\hspace{2.4mm}\tiny{$-$0.278}}}$ & $4.369 \pm 0.063 ^{\raisebox{-0.2ex}[0ex][-1ex]{\hspace{2.5mm}\tiny{$+$0.241}}}_{\raisebox{0.1ex}{\hspace{2.4mm}\tiny{$-$0.275}}}$ \\
138.1--166.9 & 151.6 & $4.376 \pm 0.202 ^{\raisebox{-0.2ex}[0ex][-1ex]{\hspace{2.5mm}\tiny{$+$0.290}}}_{\raisebox{0.1ex}{\hspace{2.4mm}\tiny{$-$0.316}}}$ & $4.510 \pm 0.078 ^{\raisebox{-0.2ex}[0ex][-1ex]{\hspace{2.5mm}\tiny{$+$0.296}}}_{\raisebox{0.1ex}{\hspace{2.4mm}\tiny{$-$0.325}}}$ \\
\enddata